\documentclass[12pt]{article}

\usepackage{subcaption} 

\usepackage{amsmath,amssymb,amsfonts}
\usepackage{graphicx}
\usepackage{caption}
\usepackage{subcaption}
\usepackage{authblk}    
\usepackage{geometry}
\usepackage[numbers, round]{natbib}
\usepackage{booktabs}
\usepackage{url}
\usepackage{hyperref}

\newcommand{\appropto}{\mathrel{\vcenter{
  \offinterlineskip\halign{\hfil$##$\cr
\propto\cr\noalign{\kern2pt}\sim\cr\noalign{\kern-2pt}}}}}

\newlength{\figwidth}
\usepackage{placeins}

\usepackage{threeparttable}

\title{Flapping Wings Amplify Pitch Stability: Insights from a Robotic Bird}

\author[a,1]{Rónán Gissler}
\author[a]{Kenneth S. Breuer}

\affil[a]{Center for Fluid Mechanics, School of Engineering, Brown University, Providence, RI, 02914}
\affil[1]{\small To whom correspondence should be addressed. E-mail: ronan\_gissler@brown.edu}
\date{\today} 

\begin{document}

\maketitle

\begin{abstract}
Using a flapping robot in a wind tunnel, we show that flapping faster amplifies existing longitudinal static stability (focusing on the pitch stiffness) and can even make an unstable flier stable. We show that stability for a flapper is not just a function of the static margin, but also the Strouhal number (St). Experimental data from measurements over a wide range of frequencies and wind speeds show good agreement with a quasi-steady blade-element (QSBE) model and a low-order approximation of the QSBE model. 
The increase in pitch stiffness at higher $\text{St}$ can primarily be explained by the increase in the mean effective wind speed. If wingbeat amplitude was allowed to vary, the model suggests that the pitch stiffness would increase with amplitude at high St but decrease with amplitude at low St. Despite using simplified wingbeat kinematics and a restricted analysis of stability, these results provide insight into how altering wingbeat kinematics can affect the passive stability of flying animals and ornithopters.
\end{abstract}

Keywords: Longitudinal $|$ Stability $|$ Flapping $|$ Dynamics $|$ Birds $|$ Bats

\section*{Introduction}
To understand the dynamics of animals and vehicles in flight, we must first understand their stability and maneuverability. While the stability of fixed-wing aircraft and gliding animal flight is well-understood \citep[e.g.][]{Cook2007}, the impact of flapping wings on flight stability remains unclear due to numerous challenges. These challenges include the role of several new kinematic parameters (flapping frequency, amplitude, stroke plane, downstroke ratio, etc) as well as dynamic considerations such as the  relative mass of the wings to the body and the challenges associated with generation and prediction of unsteady aerodynamic forces.  

In the study of flight stability, stability is typically separated into two components: static and dynamic stability \cite{Cook2007,Nelson1998}. When examining static stability, we are interested in whether the \emph{initial} response of the system returns towards equilibrium following a small perturbation. In contrast, when examining dynamic stability, we are interested in whether the system \emph{eventually} returns to equilibrium some time after experiencing a disturbance. In this sense, static stability represents a necessary prerequisite for dynamic stability. 

The most important mode of instability during flight is longitudinal instability in which the system undergoes pitch rotation. Therefore, the analysis of static longitudinal (or pitch) stability is often the starting point for flight stability analysis, and will be the focus of the current work. To analyze longitudinal static stability, we evaluate $\partial M / \partial \alpha$, where $M$ is the pitch moment about the center of mass (CoM) and $\alpha$ is the angle of attack with respect to the oncoming air stream. In order for the system to be statically stable, there must be a restorative force due to ``pitch stiffness'': $\partial M / \partial \alpha < 0$.
Note that $\partial M / \partial \alpha < 0$ is called a \emph{positive} pitch stiffness since it is a restorative force (Hooke's law). Both ``pitch stiffness" and ``pitch stability slope" are used in the text, although they are opposite in sign.

Many species of insects, birds, and bats exhibit high degrees of flight maneuverability, which is usually thought of as being in tension with flight stability \cite{Cook2007}. Complicating the issue further is the role of feedback and control. While the animal's underlying flight ``hardware'' (wing-body geometry and kinematics) might be statically and/or dynamically unstable in the absence of control, animals have extensive sensory networks (vestibular, hairs, feathers, visual, etc) that --- coupled with feedback (neural) control --- can stabilize an otherwise unstable flight system.  The role of unsteady wing motion on passive stability is thus of pressing importance for understanding the relative roles of mechanics versus neural control on stability, and how animals cope in the face of reduced sensory input (due to injury, darkness, etc).  Furthermore, passive stability is thought to have played an important role in the evolution of animal flight  as it would have allowed prehistoric animals to become airborne prior to the development of advanced neural abilities for flight control \cite{MaynardSmith1952, Evangelista2014}. Lastly, flapping wing bio-inspired engineered vehicles may offer some benefits over fixed-wing or rotor systems, such as easy transition between regimes of high efficiency and high maneuverability, reduced noise, or hide-in-plain-sight capabilities. Therefore, understanding the impact of flapping wings on flight stability has important consequences for advanced vehicle designs. 

\section*{Past Work}
The quantitative study of longitudinal (or pitch) stability in flapping flight began nearly twenty years ago with \citet{Taylor2002} who used a quasi-steady blade element (QSBE) model. QSBE models --- which have been used over seventy years to analyze both fixed and flapping-wing flight \citep{Glauert1926,Weis-Fogh1956,Ellington1984} --- partition the wing into 2D airfoil strips or ``blade elements" to account for spanwise variation in effective angle of attack and effective wind speed owing to the flapping motion of the wing as well as variable wing geometry and twist. Forces and moments are calculated at each blade element using the result predicted for a two-dimensional wing section in steady flow at the same effective wind speed and effective angle of attack. The result for the full wing is obtained by integrating the contributions from all blade elements.

When analyzing longitudinal static stability using a QSBE model, \citet{Taylor2002} found that, in theory,  flapping in forward flight is not destabilizing and can even enhance stability if the flier is already stable in gliding. In a later study, \citet{Taylor2003} collected force and moment data from tethered live locusts flapping in a wind tunnel and found that $\partial M / \partial \alpha < 0$, indicating that the locust flight exhibited static stability. 
\citet{Krashanitsa2009,LeeExp2012} collected force and moment data from robotic ornithopters in a wind tunnel and also found that $\partial M / \partial \alpha < 0$, indicating that their ornithopters exhibited static stability. \citet{Krashanitsa2009} showed the stability slope --- $\partial M / \partial \alpha$ --- became more negative with increased throttle (related to flapping frequency), but they did not report any quantitative details or examine the relationship between throttle and stability slope in detail. Furthermore, in their experiments there was no direct link between throttle and flight kinematics since even for a fixed throttle the wingbeat frequency varied depending on the wind speed and body pitch angle \cite{Krashanitsa2009}.
Both of these ornithopters also exhibited dynamic stability, demonstrated by successful free flights with fixed controls \cite{Krashanitsa2009,LeeExp2012}.

\begin{figure*}[hbp]
    \centering
    \includegraphics[width=\textwidth]{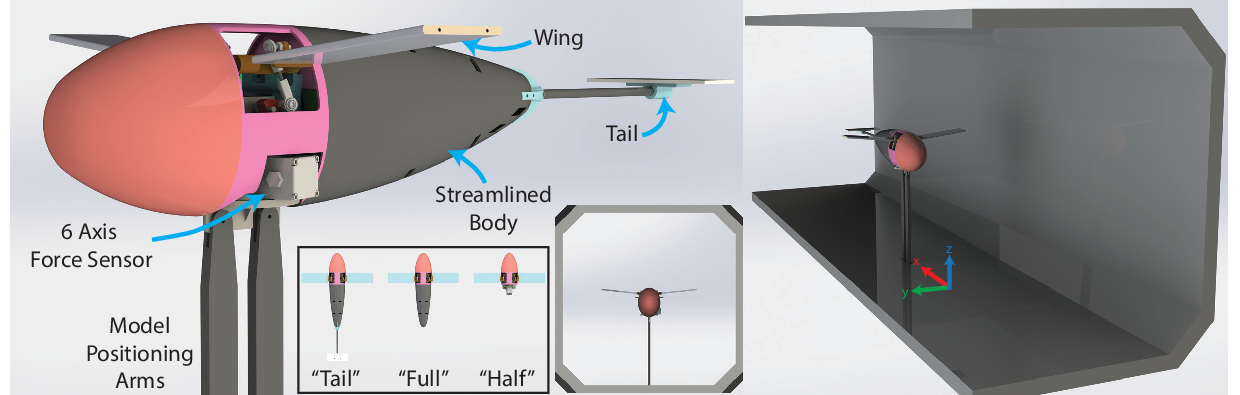}
    \caption{Experimental setup in the wind tunnel. The flapping robot has a single DoF flapping motion, confining the wings to move only in the body's transverse plane. The model positioning arms are used to adjust the pitch angle and thus the angle of attack of the robot. The three body configurations are also shown: ``Tail", ``Full", and ``Half". Unless otherwise noted, all data presented here is from the ``Half" configuration.}
    \label{fig:Flapperoo_OG}
\end{figure*}

\begin{figure}[htbp]
    \centering
    \begin{subfigure}[b]{0.49\textwidth}
        \centering
        \includegraphics[width=\textwidth]{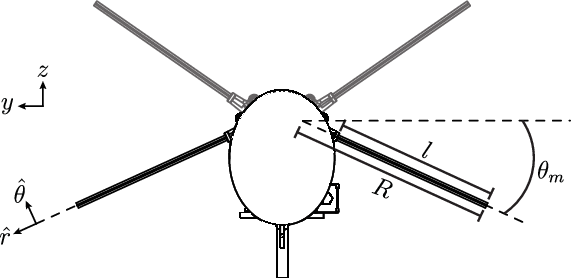}
        \caption{}
        \label{span}
    \end{subfigure}
    \begin{subfigure}[b]{0.2\textwidth}
        \centering
        \includegraphics[width=\textwidth]{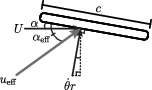}
        \caption{}
        \label{chord}
    \end{subfigure}
    \caption{(a) Diagram of the flapping wing showing geometric parameters of the wingbeat kinematics and coordinate systems. (b) Diagram of the velocities relative to the wing in a plane normal to the wingspan axis. $u_{\text{eff}}$ is the effective air velocity owing to the freestream velocity $U$ and the flapping velocity $\dot{\theta}r$. $\alpha$ is the geometric angle of attack while $\alpha_{\text{eff}}$ is the effective angle of attack.}
    \label{schematic}
\end{figure}

Despite these contributions, no experimental work to date has systematically evaluated the impact of different wingbeat kinematics on passive stability. To address this gap, we built an idealized rigid wing ornithopter with a single degree of freedom flapping wing motion (Figures~\ref{fig:Flapperoo_OG} and \ref{schematic}) and conducted wind tunnel experiments to measure the aerodynamic forces and moments over a wide range of  wingbeat frequencies, wind speeds, and body pitch angles (equivalent to the geometric angle of attack of the wings at midstroke). We compare the measurements with the predictions of a QBSE model.  Table~\ref{tab:parameters} lists the dimensions and testing parameters of the experiment. By measuring the average dimensionless pitch moment coefficient about the leading edge of the wings ($\overline{C}_{M_{LE}}$) as a function of angle of attack ($\alpha$), we evaluate how changes to wingbeat frequency and wind speed affect longitudinal static stability.

\begin{table}[htb]
\centering
\caption{Experiment Parameters}
\label{tab:parameters}
\begin{tabular}{lr}
\toprule
Wing chord $c$ & 10 cm\\
Single wing span $l$ & 25 cm\\
Full wing span & 67.5 cm\\
Wingtip to rotation axis $R$ & 31.3 cm\\
Single wing mass & 10 g\\
Wingbeat amplitude (centerline to peak) $\theta_{m}$ & $30^{\circ}$ \\
Force sensor center to LE (along x-axis) & 6.3 cm \\
\midrule
Angles of attack $\alpha$ & -16 -- 16 degrees\\
Wingbeat frequencies $f$ & 0 -- 5 Hz\\
Wind speeds $U$ & 0 -- 6 m/s\\
\midrule
Reynolds number: $\displaystyle \text{Re} = \frac{U c}{\nu} $ & 0 -- 40,000 \vspace{0.5em}\\
Strouhal number: $\displaystyle St = \frac{2 R \sin (\theta_{m}) f }{U}$ & 0 -- 0.51 \rule{0pt}{16pt} \\
\bottomrule
\end{tabular}
\end{table}

\section*{Theoretical Model}
In order to validate and interpret the experimental data and to expand the parameter space beyond the experimental configurations tested, we have used a quasi-steady blade-element (QSBE) model in which forces and moments are a function of time and spanwise position:
\begin{equation}
0 < t < \frac{1}{f}, \qquad R - l < r < R,
\end{equation}
where $t$ is time, $f$ is the wingbeat frequency, $r$ is the spanwise location of the blade element, $l$ is the  single wing span, and $R$ is the distance from the wing tip to the wing's rotational axis (Fig.~\ref{schematic}).
We assume that the wing angle, $\theta$, varies sinusoidally:
\begin{equation}
\theta = \theta_m \cos(2 \pi f t), 
\end{equation}
where the wingbeat amplitude $\theta_m = {\pi}/{6}$ in our experiments. Note that, due to the mechanical design of our model, a sinusoidal motion is not perfectly accurate (Fig.~\ref{true_vs_sin_kinematics}). However the differences are small and the results obtained using a sinusoidal approximation agree well with the experimental measurements. 

Forces and moments are calculated at each spanwise blade-element using the result predicted for steady flow over a wing segment at the same effective wind speed and effective angle of attack, resulting in a moment coefficient evaluated about the leading edge:
\begin{equation}
c_{M_{LE}} = a_0 \cos [\theta (t)] \alpha_{\text{eff}} (t,r) \left( \frac{u_{\text{eff}}(t,r)}{U} \right)^2,
\label{eqt:blade_element}
\end{equation}
\begin{equation}
a_0 = \left. \frac{\partial C_{M_{LE}}}{\partial \alpha} \right|_{f=0} ,
\end{equation}
where $a_0$ is the pitch moment coefficient slope about the leading edge of the wing when the wings are held fixed at midstroke. $a_0$ is equal to $-\pi/2$ for an ideal thin airfoil \cite{Anderson2017}, but here we obtain the value experimentally from gliding measurements of our robot (Fig.~\ref{dMda_Re}). $\alpha_{\text{eff}} (t,r)$ is the effective angle of attack, $u_{\text{eff}} (t,r)$ is the effective wind speed, and $U$ is the freestream wind speed. $u_{\text{eff}} (t,r)$ and $\alpha_{\text{eff}} (t,r)$ are derived in the SI. Although for any given instance of the experiment $u_{\text{eff}} (t,r)$ and $\alpha_{\text{eff}} (t,r)$ are only functions of time and spanwise position, they also depend on the angle of attack and Strouhal number. The Strouhal number, $St=2R \sin(\theta_m)f / U$,  is often used to characterize flapping-wing dynamics, with many flying animals exhibiting Strouhal numbers in the range $0.2 < St < 0.4$ during cruising flight \citep{TaylorSt2003}. $\cos \theta$ appears in equation \ref{eqt:blade_element} since only a portion of the moment about the wing's spanwise axis contributes to the pitch moment of the body.

To obtain the result for the entire wing, the contributions from all blade elements are integrated over the wingspan:
\begin{equation}
C_{M_{LE}} = \frac{1}{l} \int_{R-l}^R c_{M_{LE}} \; dr, 
\end{equation}
and the average moment is then evaluated over the wingbeat period:
\begin{equation}
\overline{C}_{M_{LE}} = f \int_0^{1/f} C_{M_{LE}} \; dt.
\end{equation}
Wing inertia and added mass effects are neglected as they are assumed to average to zero over a wingbeat cycle.

Combining these equations, the pitch stability slope is then:
\begin{equation}
\frac{\partial \overline{C}_{M_{LE}}}{\partial \alpha} = h \frac{\partial}{\partial \alpha} \left[ \int_0^{1/f} \cos \theta (t) \int_{R-l}^R \alpha_{\text{eff}} (t,r) u^2_{\text{eff}} (t,r) \; dr \; dt \right] , 
\label{eqt:model}
\end{equation}
where
\begin{equation}
h = a_0 \frac{f}{U^2 l} .
\label{eqt:model_constant}
\end{equation}

Equation~\ref{eqt:model} must be evaluated numerically, but by expanding the integrand for small values of the Strouhal number and angle of attack --- retaining up to quadratic terms --- we can obtain:
\begin{align}
\frac{\partial \overline{C}_{M_{LE}}}{\partial \alpha}
\approx\; & a_0 \frac{f}{\hat{l}} \frac{\partial}{\partial \alpha}\int_0^{1/f} \cos \theta(t) \int_{1-\hat{l}}^1 \bigg[ 
\alpha + \frac{\pi}{2} \hat{r} \, St \sin (2 \pi  f t) \nonumber \\
& \times \left(3 \alpha^2 + 4 \pi \alpha \hat{r} \, St \sin (2 \pi  f t) + 2 \right) 
\bigg] \, d\hat{r} \, dt ,
\label{eqt:amp_scale_full}
\end{align}

where $\alpha$ is the geometric angle of attack, $\hat{r} = r/R$ is the normalized spanwise position, $\hat{l} = l/R$ is the ratio of span to rotation arm length, and the Strouhal number $St$ is defined using the small angle approximation for $\theta_m$.
After integrating this, we obtain a simplified QSBE model for the cycle-averaged pitch moment slope:
\begin{equation}
\frac{\partial \overline{C}_{M_{LE}}}{\partial \alpha} \approx a_0 \bigg[
J_0 (\theta_m) 
+ \frac{2\pi^2}{3} \left(\hat{l}^2 - 3\hat{l} + 3\right) \frac{J_1 (\theta_m)}{\theta_m} St^2 \bigg] ,
\label{eqt:amp_scale}
\end{equation}

where $J_i$ are Bessel functions of the first kind. 
Using  the values of $l$, $R$, and $\theta_m$ for our robot (Table \ref{tab:parameters}), equation \ref{eqt:amp_scale} simplifies to:
\begin{equation}
    \frac{\partial \overline{C}_{M_{LE}}}{\partial \alpha} \approx a_0 (0.93 + 3.95 St^2) .
\label{eqt:red_model}
\end{equation}

\subsection*{Scaling}

Examining equation \ref{eqt:amp_scale}, in the fixed-wing limit ($\theta_m = 0$ and $St = 0$), we recover the fixed-wing pitch slope $a_0$. Therefore, the expression in brackets represents the modification of the pitch stiffness due to flapping. In this way, the simplified QSBE model provides more intuition into the underlying mechanics than that provided by the full form of the model.

\begin{figure}[htbp]
        \centering
        \includegraphics[width=0.5\columnwidth]{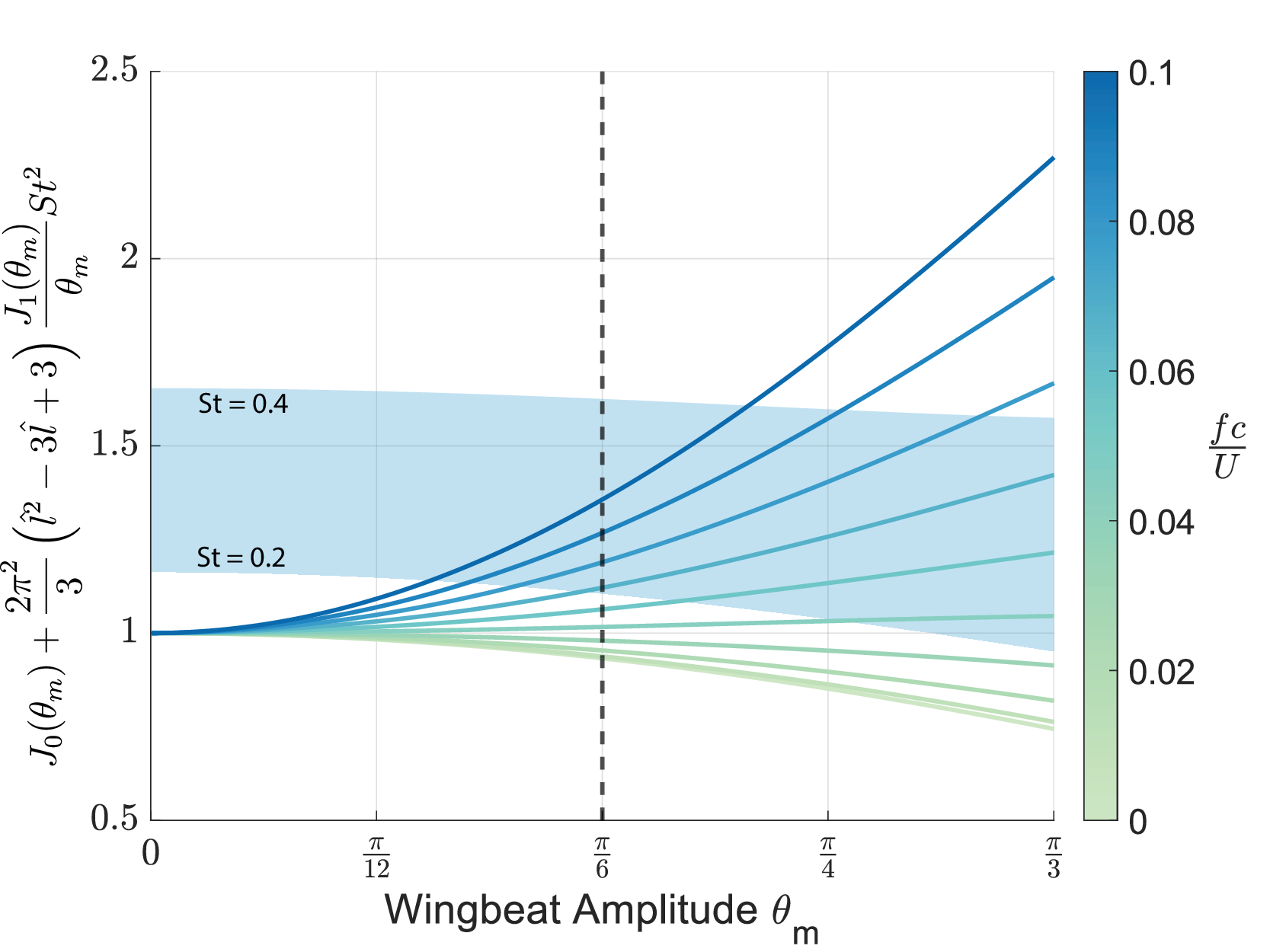}
    \caption{The multiplicative factor representing the contribution of flapping to pitch stiffness as a function of Strouhal number (equation \ref{eqt:amp_scale}). The commonly quoted biologically relevant Strouhal range of $0.2 < St < 0.4$ \cite{TaylorSt2003} is highlighted in blue. The fixed wingbeat amplitude measured in this experiment is marked by the vertical dashed line. As $\theta_m$ approaches zero in the fixed-wing limit, the pitch stiffness returns to its fixed wing value, so the multiplicative factor approaches one.}
    \label{fig:stiffness_scaling}
\end{figure}

The magnitude of the bracketed term is shown in Fig.~\ref{fig:stiffness_scaling} for a range of Strouhal numbers. The commonly quoted biologically relevant Strouhal range of $0.2 < St < 0.4$ \cite{TaylorSt2003} is highlighted in blue. The fixed wingbeat amplitude used in the current experiment is marked by the vertical dashed line. We see that for typical $St$ (blue region), the predicted pitch stiffness is greater than the fixed wing stiffness (i.e. the factor is greater than one) though the magnitude diminishes at higher amplitudes. However, there are a few cases of low-frequency, high-amplitude flapping (outside the biologically relevant range of Strouhal numbers) which predict a decrease in pitch stiffness compared to stationary wings. Additionally, the simplified QSBE model predicts that the pitch stiffness should always increase if the ratio of wingbeat frequency to wind speed is increased, while the pitch stiffness generally increases if the wingbeat amplitude is increased, though at low frequency, increasing wingbeat amplitude is predicted to decrease the pitch stiffness.

Examining the intersection of the dashed line and the region highlighted in blue, we see that for the typical $St$ range and the wingbeat amplitude used here, we expect gains in the pitch stiffness between $11\%$ and $63\%$ compared to fixed wings. Note that these gains are evaluated with the center-of-mass located at the leading edge of the wing. The sensitivity of pitch stiffness to center-of-mass position will be discussed in more detail later.

\section*{Experimental Results}
\begin{figure}[htbp]
    \centering
    \includegraphics[width=0.5\columnwidth]{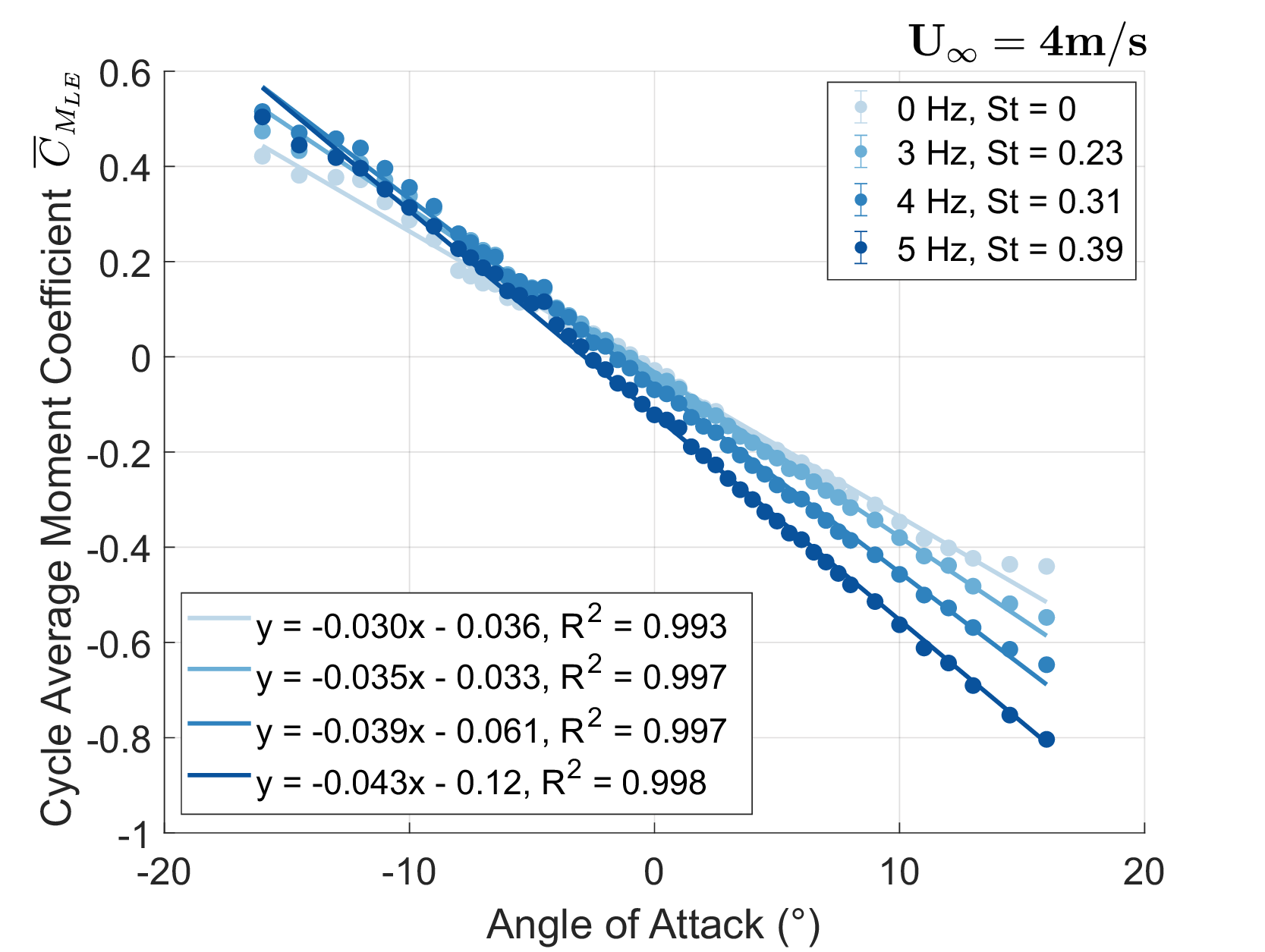}
    \caption{Typical portrait of static stability: cycle average pitch moment coefficient about the leading edge versus angle of attack for several wingbeat frequencies at a wind speed of 4 m/s. The standard deviation of $\overline{C}_{M_{LE}}$ across all 180 wingbeats is plotted as the error bars ($\pm 1$ SD) but they are imperceptibly small. The slope of each curve corresponds to the magnitude of the response when the system is perturbed from equilibrium. Since the slopes are negative, the response is stabilizing. As the flapping frequency is increased, it is apparent that the slope becomes more negative. In other words, flapping faster increases static stability. This plot is included to highlight the linearity of the pitch moment data over the range of angles considered.}
    \label{dMda_Hz}
\end{figure}

Fig.~\ref{dMda_Hz} presents one set of measurements of the cycle-averaged pitch moment coefficient evaluated about the leading edge, $\overline{C}_{M_{LE}}$, as a function of the angle of attack for four different flapping frequencies, all at a fixed wind speed.
It is apparent that the data over the range of $\alpha$ tested is sufficiently linear such that the stability slope $\partial \overline{C}_{M_{LE}} / \partial \alpha$ can be considered independent of $\alpha$ for this range. 
Additionally, increasing the wingbeat frequency results in a more negative stability slope or an increase in the pitch stiffness.

\begin{figure}[htbp]
        \centering
        \includegraphics[width=0.5\columnwidth]{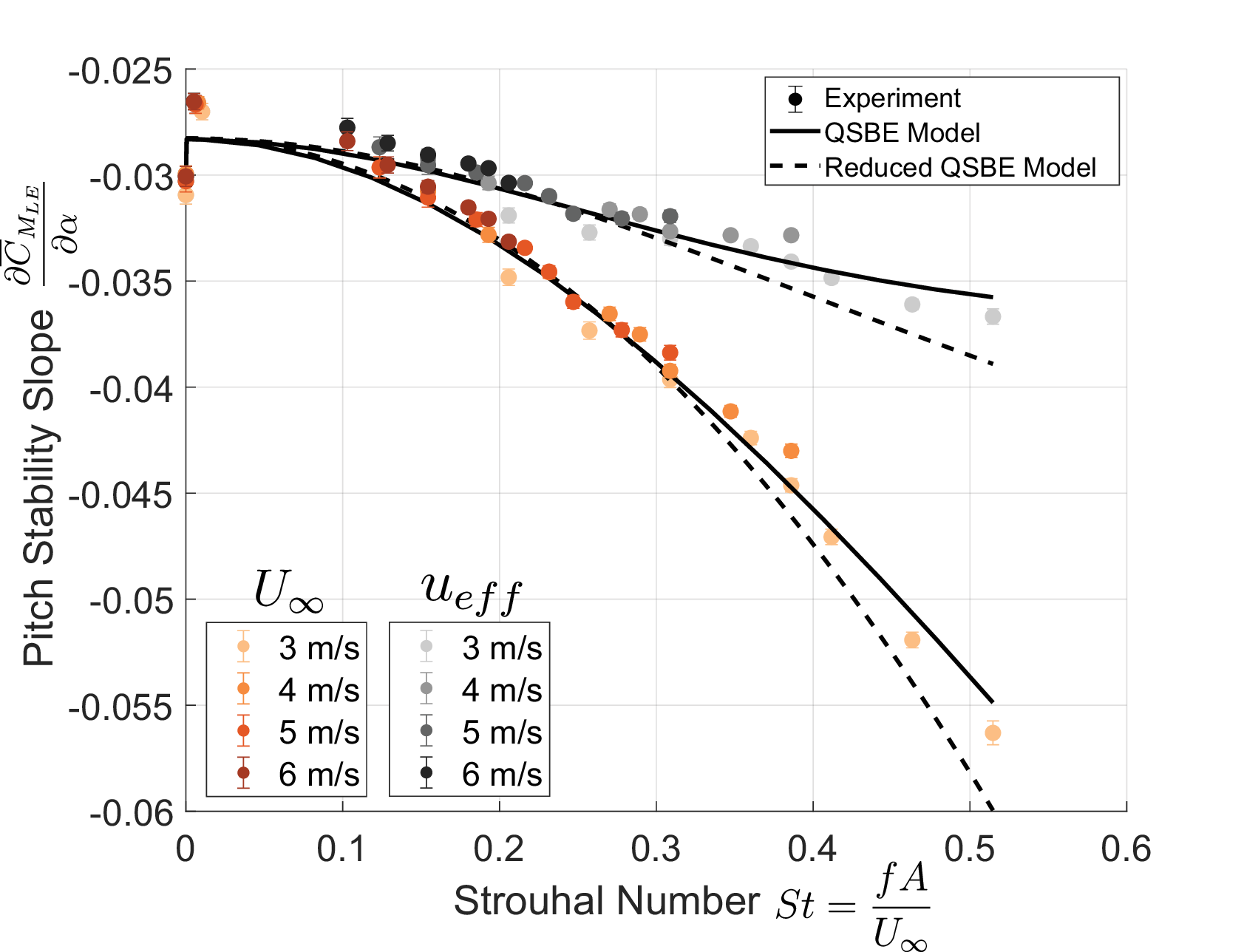}
    \caption{Pitch stability slope (inverse of pitch stiffness) as a function of Strouhal number. In shades of red are the data nondimensionalized by the freestream wind speed. In shades of gray are the data nondimensionalized by the mean effective wind speed of the wing. The standard error for the slope estimate of the linear regression is plotted as the error bars ($\pm 1$ SE). All following statements refer to the data in red. Fitting a power law to the experimental flapping data ($St \neq 0$), we obtain $\partial \overline{C}_{M_{LE}} / \partial \alpha = -0.087 St^{1.66} - 0.027$. Fitting a power law to the QSBE flapping data, we obtain $\partial \overline{C}_{M_{LE}} / \partial \alpha = -0.082 St^{1.77} - 0.028$. The reduced QSBE model is given by $\partial \overline{C}_{M_{LE}} / \partial \alpha = -0.12 St^2 - 0.028$ (Equation \ref{eqt:red_model}).}
    \label{fig:stability_slopes}
\end{figure}

In Fig.~\ref{fig:stability_slopes}, the stability slope is plotted for every wind speed and wingbeat frequency combination tested. 
Each dot on Fig.~\ref{fig:stability_slopes} is obtained from the linear regression of measurements across 49 different angles of attack.
The solid lines show the QSBE model prediction (equation \ref{eqt:model}) while the dashed lines show the prediction of the reduced form of the QSBE model (equation \ref{eqt:red_model}).
The data in shades of red are nondimensionalized by the freestream wind speed ($U$) while the data in shades of gray are nondimensionalized by the mean effective wind speed ($u_{\text{eff}}$), given by:

\begin{align}
    \overline{u}_{\text{eff}} &= \frac{f}{l} \int_0^{1/f} \int_{R-l}^R u_{\text{eff}} (t,r) \; dr \; dt \\
    &= U \sqrt{1 + \frac{\pi^2}{6} \left(\hat{l}^2 - 3\hat{l} + 3\right) St^2}.
\label{eqt:eff_vel}
\end{align}

Note the similarity between equation \ref{eqt:eff_vel} and equation \ref{eqt:amp_scale} and how the effective wind speed grows with Strouhal number.

We begin by focusing on the data in shades of red as it uses the typical nondimensionalization.
Most importantly, there is an excellent collapse of the data with Strouhal number showing an increase in pitch stiffness as $St$ increases.
This provides experimental support to the same finding of earlier theoretical works \cite{Taylor2002, Taha2015}.
There is also generally good agreement between the measurements and the QSBE model, and even the simplified (low $St$) model.

Examining the region at $St = 0$, the discontinuous jump from $St = 0$ to the first nonzero $St$ marks a fundamental difference between gliding and flapping.
This jump is recorded in the measurements and also predicted by the model (albeit with a smaller magnitude).
Examining the terms of the model, it can be explained by the $\cos\theta(t)$ term in Equation \ref{eqt:model}.
$\theta(t) = 0$ for gliding while $\theta(t) = \theta_m \cos (2 \pi f t)$ for flapping, and thus the mean value of $\cos\theta(t)$ will change discontinuously between gliding and flapping.
Looking at equation \ref{eqt:model}, transitioning to flapping would then cause the pitch stiffness to decrease.
As further confirmation, without $\cos\theta(t)$ the discontinuity disappears from the model (Fig.~\ref{model_variants}).

Since the wingbeat amplitude is fixed in our experiments, the $\cos\theta(t)$ term does not change further with $St$.
When increasing $St$, other mechanisms come into play to alter the pitch stiffness, namely the increase in mean effective wind speed.
Increasing the effective wind speed increases the dynamic pressure which increases the magnitude of the moment at all angles of attack, thereby amplifying the slope \cite{Taylor2002}.
The effect of the effective wind speed outweighs that of the effective angle of attack as can be seen when removing each term from the model (Fig.~\ref{model_variants}). 

In Fig.~\ref{fig:stability_slopes}, we show (in grayscale) the same data, now rescaled with a dynamic pressure that uses the mean effective wind speed ($\overline{u}_{\text{eff}}$, which is always higher than the freestream speed).
This rescaling reduces the variation of the dimensionless pitch stiffness with $St$ dramatically, but even so the data still show a dependence on $St$.
If the relationship between pitch stiffness and $St$ depended only on how $St$ changes $\overline{u}_{\text{eff}}$, then we would expect the grayscale data in Fig.~\ref{fig:stability_slopes} to appear as a flat line. 

While we've shown the pitch stiffness scales with $St$, the reduced QSBE model from Equation \ref{eqt:amp_scale} suggests that the dependence of the pitch stiffness on wingbeat amplitude is more complicated.
For low $St$, increasing the amplitude would decrease the stiffness, while for high $St$ it would increase the stiffness (Fig.~\ref{amp_model}).
Since the wingbeat amplitude was fixed in these experiments, we could not verify this prediction of the model.
Nonetheless, the convenient and popular QSBE model performs well in predicting pitch stiffness.
This is encouraging, particularly given the large range of effective angles of attack --- up to 60 degrees at the wingtip for the highest wingbeat frequency and slowest wind speed combination (Fig.~\ref{eff_AoA}).

\subsection*{Aerodynamic Center, Center of Mass and Static Margin}

\begin{figure*}[htbp]
    \centering
    \includegraphics[width=0.9\textwidth]{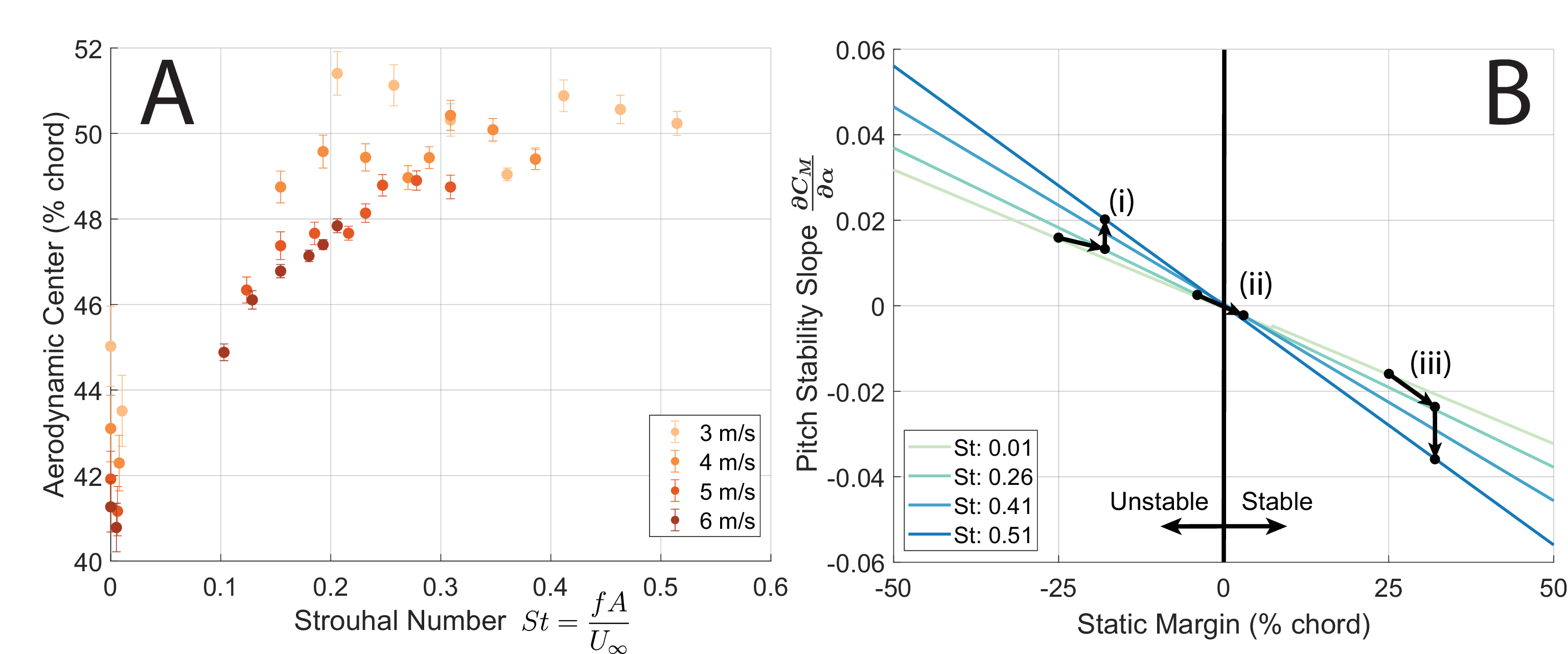}
    \caption{Static stability as a function of center of mass (CoM) location. For a given CoM, static stability increases with Strouhal number by (a) shifting the aerodynamic center (AC) further rearward thereby increasing the static margin and (b) amplification for a given static margin. The black arrows on (b) depict how the pitch stability slope would change for a given CoM as $St$ increased for three cases. Note that (a) implies that a slightly unstable configuration could be made stable by increasing the Strouhal number; this case is depicted by the black arrow marked (ii) in (b). The error bars in (a) are given by how far the position about which the moments are considered must be shifted from the AC such that the slope equals the standard error of the slope estimate at the AC.}
    \label{grand_stability}
\end{figure*}

Although we have thus far defined the pitch moment coefficient as the moment about the leading edge of the wing, it can be referenced relative to any point (see equation \ref{eqt:mom_shift} in \nameref{sec:methods}). In particular, the ``neutral point'' (NP) is defined in aerodynamic theory as the location where the pitch moment is independent of angle of attack (i.e. $\partial M / \partial \alpha = 0$). The ``aerodynamic center'' (AC) is defined in the same manner, except that the analysis applies only to the forces and moments generated by the wing in isolation. Here, the effects of the vehicle body are subtracted (as described in the \nameref{sec:methods}) so we evaluate the AC. The dependence of the AC on the Strouhal number is shown in Fig.~\ref{grand_stability}a. We see that the AC shifts slightly aft --- by about 10\% of the chord over the full range of Strouhal numbers tested --- as $St$ increases. This builds on the work of \citet{Taylor2002} who assumed the position of the AC was constant.

In a freely flying system, the pitching moment acts about the center-of-mass (CoM). A key measure of longitudinal static stability is the ``static margin'' ($\tilde{x}$), the distance between the CoM and the aerodynamic center (or neutral point), defined as $\tilde{x} = (x_{AC} - x_{CoM}) / c$ with $x = 0$ marking the leading edge of the wing and $x = c$ marking the trailing edge. When the CoM is in front of the AC --- a positive static margin --- the wing is statically stable to small perturbations in angle of attack. Since our robot is tethered, we can assume any location for the CoM, $x_\text{CoM}$, and consequently a range of possible static margins, both positive (stable) and negative (unstable).

Fig.~\ref{grand_stability}b shows the consequences of flapping for three possible choices of $x_\text{CoM}$: (i) negative $\tilde{x}$ (ii) slightly negative $\tilde{x}$ (iii) positive $\tilde{x}$.  Generally speaking, increasing $St$ amplifies the \emph{existing} stability or instability exhibited in gliding, paths (i) and (iii). However, given that increasing $St$ also shifts the AC aft (Fig.~\ref{grand_stability}A) the stability slope changes in a nonlinear fashion and can even move from unstable to stable as shown in path (ii).

\subsection*{Trim State}

While a positive static margin is necessary for longitudinal static stability, the conditions for force/moment equilibrium during flight must also be satisfied. For the longitudinal flight dynamics in level flight, this requires that the lift force equal body weight, the thrust equal the drag, and the net pitch moment be zero.

The equilibrium condition (trim state) in pitch for a wing is given by
\begin{equation}
-\tilde{x} (C_L \cos \alpha + C_D \sin \alpha) + \tilde{z} (C_D \cos \alpha - C_L \sin \alpha) + C_{M_{AC}} = 0,
\end{equation}
where $C_L$ is the lift coefficient, $C_D$ is the drag coefficient, $C_{M_{AC}}$ is the pitching moment coefficient about the aerodynamic center, and $\tilde{z} = (z_{AC} - z_{CoM}) / c$ is the vertical distance between the AC and the CoM \citep{Taylor2002, Nelson1998}. Since typically $\tilde{z} \ll \tilde{x}$ and $C_D \ll C_L$, the equilibrium condition at small angles of attack can be simplified as
\begin{equation}
C_{M_{AC}} = \tilde{x} C_L.
\end{equation}

\begin{figure*}[htb]
    \centering
    \includegraphics[width=0.9\textwidth]{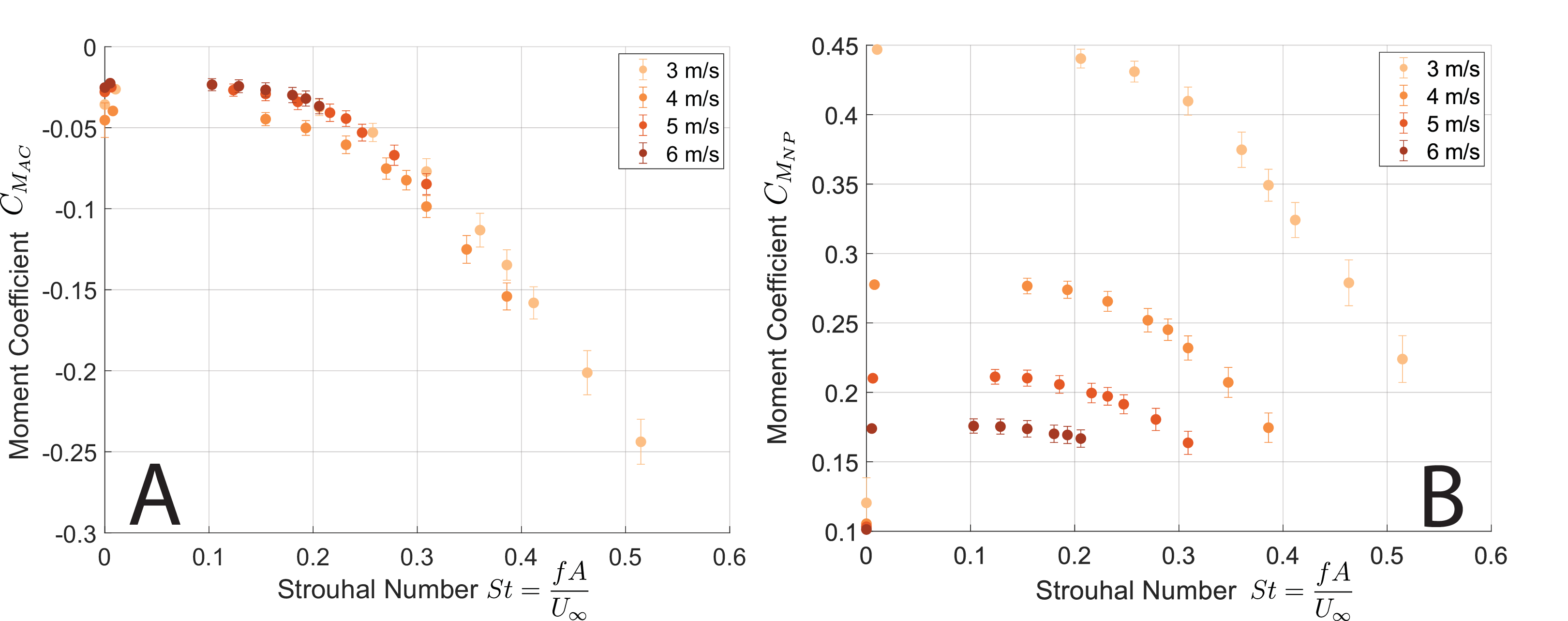}
    \caption{Pitch moment at the aerodynamic center (AC) and neutral point (NP) decreases with Strouhal number, suggesting that as static stability increases the equilibrium configuration must also change. (a) Typical body subtracted data shows no capacity for equilibrium since the moment about the AC is always negative (b) Data without the body subtraction highlight how body effects can contribute to equilibrium since the moment about the NP is positive. The error bars are calculated using SE as described in Fig.~\ref{dMda_Hz}, but propagated through Equation \ref{eqt:mom_shift} to calculate $SE(C_{M_{AC}})$.}
    \label{moment_NP}
\end{figure*}

For stable flight, the static margin $\tilde{x}$ is positive and since lift is positive, this necessitates that the moment at the AC be positive for an equilibrium condition to exist. In Fig.~\ref{moment_NP}a, we see that the moment is always negative, so there is no equilibrium state for this flier. For similar simplified kinematics using rigid wings, \citet{Taylor2002} arrived at the same conclusion theoretically and suggested that birds may produce the positive moment necessary for equilibrium through wing twist or by the use of a tail.

In Fig.~\ref{moment_NP}b, we replot our data \emph{without subtracting the forces associated with the body} and see that the moment at the NP (no longer the AC, since we include body effects) is always positive, suggesting that an equilibrium condition in pitch exists if body effects are included. Including tail effects similarly results in positive moments at the NP (Fig.~\ref{moment_NP_no_sub_tail}). 

While static stability increases with Strouhal number (Fig.~\ref{grand_stability}b), the flier must compensate in other ways to maintain trim since $\tilde{x}$ (Fig.~\ref{grand_stability}a), $C_{M_{AC}}$ (Fig.~\ref{moment_NP}a), and $C_L$ will also vary with Strouhal number. \citet{Lee2010} similarly showed that as wingbeat frequency is increased in a free flier, other flight parameters must change to maintain trim. For their flier this was achieved through a decrease in mean pitch angle and an increase in mean flight speed. 

\section*{Biological Implications}

Our experiments use a highly simplified model of animal flight and provide a limited analysis of stability (static stability is not true stability, a stiff system is not necessarily dynamically stable); given this, it is fair to ask whether real animals would increase their wingbeat frequency, wingbeat amplitude, or flight speed as a method for increasing stability.
The response of free-flying animals to perturbations has been studied in a number of experiments \citep{Ravi2015, Quinn2025, Ortega-Jimenez2014, ManuelOrtega-Jimenez2013, Badger2019, Boerma2019, Hedrick2024, Beatus2015, Ristroph2010}. However, since an animal's response to a perturbation is a combination of open-loop (passive) flight dynamics and closed-loop (active) neural control, it is impossible to definitively isolate the contribution of passive stability from these experiments.

Closed-loop control depends on sensory feedback and the magnitude of control inputs. As these inputs are restricted, the animal approaches an open-loop state. For example, it is much easier to observe the passive stability of a bicycle when riding hands free since the magnitude of control inputs are dramatically reduced \cite{Jones2006}. Similarly, animal flight can be pushed towards a more open-loop state by inhibiting some of the sensory feedback pathways.

\citet{Boublil2024} and \citet{Quinn2025} performed experiments characterizing the flight performance of bats with the sensory hairs covering the ventral surface of their wings removed (depilation treatment). Although the detailed function of these sensory hairs remains unclear, they are thought to play an important role in flight control. In both of these studies, when the sensory hairs were removed, bats increased their flight speed and increased wingbeat frequency --- a  response consistent with our results. \citet{Cheney2022} inhibited the contraction of wing membrane muscles (plagiopatagiales proprii) of bats using Botox, which prevented the animals from controlling the wing membrane camber during flight. They reported that after treatment the bats (i) would not fly at lower speeds and (ii) used a larger wingbeat amplitude when flying at higher speeds.

One possible explanation for the behavior of these bats is that the absence of sensory feedback (in the depilated group) or control authority (in the Botoxed group) forced the animals to rely more heavily on passive stabilizing mechanisms to maintain controlled flight. If variability of wingbeat kinematics is considered a proxy for the degree of active control, then additional evidence is lent to these experiments representing a more open-loop type of flight control since a number of wingbeat variables in \cite{Quinn2025}, wingbeat frequency in \cite{Boublil2024}, and wingbeat amplitude in \cite{Cheney2022} showed a reduced standard deviation after treatment.

\section*{Concluding Comments}

Testing a flapping-wing robot over a range of wingbeat frequencies and flight speeds, and with the guidance of a simple theoretical model, we have shown that increasing the Strouhal number primarily acts to amplify the existing longitudinal static stability or instability, though it can also make an unstable flier stable. 
We extended the development of the quasi-steady blade-element (QSBE) model with a simplified analytical expression that predicts the pitch stiffness is well described by $St$ alone at high $St$, but at lower $St$ the wingbeat amplitude can oppose the $St$ scaling.  
The QSBE model in its full and reduced forms agrees well with the experimental measurements of the pitch stiffness. As $St$ is changed to alter stability, the flier will need to adjust other kinematic parameters to maintain trim since the forces and moments are also functions of $St$. Our experimental data shows that the position of the aerodynamic center $x_{AC}$ and the moment about the aerodynamic center $C_{M_{AC}}$ are Strouhal dependent, something the QSBE model alone does not predict.

Of course the wing geometries and wingbeat kinematics of animals are much more complex than the rectangular wings and single degree-of-freedom kinematics explored here. Additional work is required to evaluate the impact of other important kinematics variables such as wing pitching, twist, wing extension \cite{Weston2026}, downstroke-upstroke ratio, etc. Likewise, the impact of the tail must be incorporated. It is generally the case with traditional fixed-wing aircraft that the wing and body act to destabilize the aircraft longitudinally such that a plane without a tail would always be unstable \citep{Nelson1998}. Animals are also known to use their tails extensively during unsteady flight maneuvers. 

Longitudinal flight stability is much more than pitch stiffness (static stability). A stiff system has stronger restorative responses to disturbances but stiffness does not necessitate dynamic stability. Studies evaluating dynamic stability of flapping in forward flight, have found it to be both unstable \citep{Taylor2003, Taylor2005, Xiong2008, Ducci2021} and stable \citep{Lee2010, Mwongera2012, Ducci2022} under certain conditions. Parameters that were found to impact stability include wing mounting position along the body length \citep{Mwongera2012, Ducci2021}, wingbeat frequency \citep{Mwongera2012}, wingbeat amplitude (marginal effect) \citep{Ducci2021, Ducci2022}, wing flexibility \citep{LeeExp2012}, and tail spread \cite{Ducci2022}. Experimental work evaluating dynamic stability in flapping flight (either biological or robotic) is sparse. Additional work should focus on measuring stability derivatives associated with pitching rate, which can then be used to evaluate the growth of perturbations. Alternatively, a cyber-physical system (CPS) could be used to experimentally evaluate dynamic stability by tracking the system's response to different perturbations. A CPS would have the additional benefit of allowing the evaluation of large perturbations, a limitation of linear stability analysis.

\section*{Materials and Methods}
\label{sec:methods}

The flapping mechanism consists of a right angle bevel gearbox driven by a stepper motor. The two output shafts of the gearbox each rotate a small crank link connected to a wing in a four bar linkage type arrangement. The mechanism is covered by a 3D-printed shell to provide a streamlined profile. The rectangular wings (10 x 25 cm) have no camber and are symmetric from leading to trailing edges. They have a rigid construction consisting of a balsa wood and carbon fiber tube frame wrapped in tissue paper. Force and moment measurements were collected using a six axis force/torque sensor (ATI Gamma, SI-65-5).

The exeriments were conducted in the Brown University Animal Flight and Aeromechanics wind tunnel \citep{Breuer2022}. The wind tunnel test section has a cross section of 1.2 x 1.2 meters such that the wingtips are approximately 0.3 meters from the ceiling at their closest point and 0.3 meters from the walls at their closest point (Fig.~\ref{fig:Flapperoo_OG}). Model positioning arms rotate the robot in the pitch direction to change the angle of attack of the entire system. 

Forces and moments are measured for different angles of attack, wingbeat frequencies (including a quasi-steady case at 0.1 Hz), and wind speeds (Table \ref{tab:parameters}). 
Data is recorded from the force and moment sensor using a Analog-to-Digital converter (National Instruments USB-6341) at a sampling rate of 9000 Hz. Data is collected over 180 wingbeats for the flapping cases or a period of ten seconds for the gliding cases. To account for thermal drift of the force and moment sensor, the sensor is tared between each measurement.

To isolate the contribution of the flapping wings, we remove the effect of the body by performing a ``body subtraction" wherein every experiment was replicated without the wings attached and the forces and moments were subtracted from the measurements taken with the wings attached. In addition to removing the aerodynamic effects of the body, this subtraction negates inertial effects due to gearbox vibration and shoulder movements, both of which are highly repeatable. The remaining forces and moments after this subtraction are then primarily due to the aerodynamics and inertia of the wings.

Although the effect of the body was not the primary interest of this study (particularly because of its disproportionately large size required to accommodate the flapping mechanism), we replicated all experiments with three different body configurations: body with tail, full body, and half body (Fig.~\ref{fig:Flapperoo_OG}). Considering all the parameters that were varied --- angle of attack, wingbeat frequency, wind speed, body configurations, with and without wings --- 14,112 unique trials were recorded in total. Unless otherwise noted, all data presented is from the ``half" body configuration for which the wing-body interaction was expected to be minimized. The same trends were observed for all body configurations (Fig.~\ref{slopes_tail} compares ``half" and ``tail").

In order for these tethered experiments to be an adequate model of free flight, we assume that --- if the body was free to pitch --- the rotation of the body over the course of a wingbeat would be sufficiently small so that we could ignore any contributions to the average pitch moment from the flapping-induced pitching motion of the body (see note on quasi-static assumption in SI).

The location of the aerodynamic center (AC) and neutral point (NP) were calculated through an iterative procedure by shifting the location of the pitch moment until the slope of the pitch moment ($M$) vs. angle of attack ($\alpha$) curve approached zero (thresholded at $10^{-6} Nm$). The pitch moment was shifted using
\begin{equation}
N = L \cos \alpha + D \sin \alpha,
\qquad
M' = M + N x.
\label{eqt:mom_shift}
\end{equation}

\section*{Acknowledgments}
Many individuals helped construct the experimental setup and run experiments including Oliver Sand, Xiaozhou Fan, Ben Lyons, Sakthi Swarrup, Rehaan Irani, and Daniel Marella. Sharon Swartz and Brooke Quinn provided helpful feedback on the biological implications of these experiments with robots. This work was supported by the National Science Foundation (Award 1930924) and the Office of Naval Research (Award N00014-21-1-2816).

\bibliographystyle{unsrtnat}
\bibliography{export}

@article{Taylor2002,
   author = {G. K. Taylor and A. L.R. Thomas},
   doi = {10.1006/jtbi.2001.2470},
   issn = {00225193},
   issue = {3},
   journal = {Journal of Theoretical Biology},
   pages = {351-370},
   pmid = {11846595},
   title = {Animal flight dynamics II. Longitudinal stability in flapping flight},
   volume = {214},
   year = {2002}
}

@book{Anderson2017,
   author = {John Anderson},
   city = {New York},
   edition = {6},
   publisher = {McGraw-Hill Education},
   title = {Fundamentals of Aerodynamics},
   year = {2017}
}

@article{ManuelOrtega-Jimenez2013,
   author = {Victor Manuel Ortega-Jimenez and Jeremy S. M. Greeter and Rajat Mittal and Tyson L. Hedrick},
   doi = {10.1242/jeb.089672},
   issue = {24},
   journal = {Journal of Experimental Biology},
   pages = {4567-4579},
   title = {Hawkmoth flight stability in turbulent vortex streets},
   volume = {216},
   year = {2013}
}

@inproceedings{Mwongera2012,
   author = {Victor M. Mwongera and Mark H. Lowenberg},
   doi = {10.2514/6.2012-4407},
   isbn = {9781624101847},
   booktitle = {AIAA Atmospheric Flight Mechanics Conference 2012},
   title = {Bifurcation analysis of a flapping wing MAV in longitudinal flight},
   year = {2012}
}

@article{Ducci2021,
   author = {Gianmarco Ducci and Victor Colognesi and Gennaro Vitucci and Philippe Chatelain and Renaud Ronsse},
   doi = {10.1007/s00332-021-09698-1},
   issn = {14321467},
   issue = {2},
   journal = {Journal of Nonlinear Science},
   month = {4},
   publisher = {Springer},
   title = {Stability and Sensitivity Analysis of Bird Flapping Flight},
   volume = {31},
   year = {2021}
}

@article{Xiong2008,
   author = {Yan Xiong and Mao Sun},
   doi = {10.1007/s10409-007-0121-2},
   issn = {05677718},
   issue = {1},
   journal = {Acta Mechanica Sinica/Lixue Xuebao},
   month = {2},
   pages = {25-36},
   title = {Dynamic flight stability of a bumblebee in forward flight},
   volume = {24},
   year = {2008}
}

@article{Boerma2019,
   author = {David B. Boerma and Kenneth S. Breuer and Tim L. Treskatis and Sharon M. Swartz},
   doi = {10.1242/jeb.204255},
   issn = {00220949},
   issue = {20},
   journal = {Journal of Experimental Biology},
   pmid = {31537651},
   publisher = {Company of Biologists Ltd},
   title = {Wings as inertial appendages: How bats recover from aerial stumbles},
   volume = {222},
   year = {2019}
}

@article{Badger2019,
   author = {Marc A. Badger and Hao Wang and Robert Dudley},
   doi = {10.1242/jeb.176263},
   issn = {00220949},
   issue = {3},
   journal = {Journal of Experimental Biology},
   month = {2},
   pmid = {30718291},
   publisher = {Company of Biologists Ltd},
   title = {Avoiding topsy-turvy: How Anna’s hummingbirds (Calypte anna) fly through upward gusts},
   volume = {222},
   year = {2019}
}

@article{Ravi2015,
   author = {Sridhar Ravi and James D. Crall and Lucas McNeilly and Susan F. Gagliardi and Andrew A. Biewener and Stacey A. Combes},
   doi = {10.1242/jeb.114553},
   issn = {14779145},
   issue = {9},
   journal = {Journal of Experimental Biology},
   month = {5},
   pages = {1444-1452},
   pmid = {25767146},
   publisher = {Company of Biologists Ltd},
   title = {Hummingbird flight stability and control in freestream turbulent winds},
   volume = {218},
   year = {2015}
}

@article{Ortega-Jimenez2014,
   author = {Victor M. Ortega-Jimenez and Nir Sapir and Marta Wolf and Evan A. Variano and Robert Dudley},
   doi = {10.1098/rspb.2014.0180},
   issn = {14712954},
   issue = {1783},
   journal = {Proceedings of the Royal Society B: Biological Sciences},
   month = {3},
   pmid = {24671978},
   publisher = {Royal Society},
   title = {Into turbulent air: Size-dependent effects of von Kármán vortex streets on hummingbird flight kinematics and energetics},
   volume = {281},
   year = {2014}
}

@article{Dietl2008,
   author = {John M. Dietl and Ephrahim Garcia},
   doi = {10.2514/1.33561},
   issn = {15333884},
   issue = {4},
   journal = {Journal of Guidance, Control, and Dynamics},
   pages = {1157-1162},
   publisher = {American Institute of Aeronautics and Astronautics Inc.},
   title = {Stability in ornithopter longitudinal flight dynamics},
   volume = {31},
   year = {2008}
}

@article{Ducci2022,
   author = {Gianmarco Ducci and Gennaro Vitucci and Philippe Chatelain and Renaud Ronsse},
   doi = {10.1038/s41598-022-27179-7},
   issn = {20452322},
   issue = {1},
   journal = {Scientific Reports},
   month = {12},
   pmid = {36587181},
   publisher = {Nature Research},
   title = {On the role of tail in stability and energetic cost of bird flapping flight},
   volume = {12},
   year = {2022}
}

@article{Taylor2003,
   author = {Graham K. Taylor and Adrian L.R. Thomas},
   doi = {10.1242/jeb.00501},
   issn = {00220949},
   issue = {16},
   journal = {Journal of Experimental Biology},
   month = {8},
   pages = {2803-2829},
   pmid = {12847126},
   title = {Dynamic flight stability in the desert locust Schistocerca gregaria},
   volume = {206},
   year = {2003}
}

@article{LeeExp2012,
   author = {Jun Seong Lee and Jae Hung Han},
   doi = {10.1088/0964-1726/21/9/094023},
   issn = {09641726},
   issue = {9},
   journal = {Smart Materials and Structures},
   month = {9},
   title = {Experimental study on the flight dynamics of a bioinspired ornithopter: Free flight testing and wind tunnel testing},
   volume = {21},
   year = {2012}
}

@article{Evangelista2014,
   author = {Dennis Evangelista and Sharlene Cam and Tony Huynh and Austin Kwong and Homayun Mehrabani and Kyle Tse and Robert Dudley},
   doi = {10.7717/peerj.632},
   issn = {21678359},
   issue = {1},
   journal = {PeerJ},
   pmid = {25337460},
   publisher = {PeerJ Inc.},
   title = {Shifts in stability and control effectiveness during evolution of Paraves support aerial maneuvering hypotheses for flight origins},
   volume = {2014},
   year = {2014}
}

@article{Taha2015,
   author = {Haithem E. Taha and Sevak Tahmasian and Craig A. Woolsey and Ali H. Nayfeh and Muhammad R. Hajj},
   doi = {10.1088/1748-3190/10/1/016002},
   issn = {17483190},
   issue = {1},
   journal = {Bioinspiration and Biomimetics},
   month = {2},
   pmid = {25561166},
   publisher = {Institute of Physics Publishing},
   title = {The need for higher-order averaging in the stability analysis of hovering, flapping-wing flight},
   volume = {10},
   year = {2015}
}

@article{Krashanitsa2009,
   author = {Roman Y Krashanitsa and Dmitro Silin and Sergey V Shkarayev and Gregg Abate},
   journal = {International Journal of Micro Air Vehicles},
   title = {Flight Dynamics of a Flapping-Wing Air Vehicle},
   volume = {1},
   year = {2009}
}

@inproceedings{Lee2010,
   author = {Jun Seong Lee and Joong Kwan Kim and Dae Kwan Kim and Jae Hung Han},
   doi = {10.2514/6.2010-8237},
   isbn = {9781624101519},
   booktitle = {AIAA Atmospheric Flight Mechanics Conference 2010},
   publisher = {American Institute of Aeronautics and Astronautics Inc.},
   title = {Longitudinal flight dynamics of bio-inspired ornithopter considering fluid-structure interaction},
   year = {2010}
}

@article{Taylor2005,
   author = {Graham K. Taylor and Rafał Zbikowski},
   doi = {10.1098/rsif.2005.0036},
   issn = {17425662},
   issue = {3},
   journal = {Journal of the Royal Society Interface},
   pages = {197-221},
   pmid = {16849180},
   publisher = {Royal Society},
   title = {Nonlinear time-periodic models of the longitudinal flight dynamics of desert locusts Schistocerca gregaria},
   volume = {2},
   year = {2005}
}

@article{Boublil2024,
   author = {Brittney L. Boublil and Chao Yu and Grant Shewmaker and Susanne Sterbing and Cynthia F. Moss},
   doi = {10.1007/s00359-023-01682-2},
   issn = {14321351},
   issue = {5},
   journal = {Journal of Comparative Physiology A: Neuroethology, Sensory, Neural, and Behavioral Physiology},
   month = {9},
   pages = {761-770},
   pmid = {38097720},
   publisher = {Springer Science and Business Media Deutschland GmbH},
   title = {Ventral wing hairs provide tactile feedback for aerial prey capture in the big brown bat, Eptesicus fuscus},
   volume = {210},
   year = {2024}
}

@techReport{Grauer2012,
   author = {Jared Grauer and James E Hubbard},
   title = {Title of dissertation: MODELING AND SYSTEM IDENTIFICATION OF AN ORNITHOPTER FLIGHT DYNAMICS MODEL},
   year = {2012}
}

@techReport{Orlowski2011,
   author = {Christopher T Orlowski},
   title = {Flapping Wing Micro Air Vehicles: An Analysis of the Importance of the Mass of the Wings to Flight Dynamics, Stability, and Control},
   year = {2011}
}

@book{Cook2007,
   author = {Michael V. Cook},
   edition = {2},
   publisher = {Elsevier Science \& Technology},
   title = {Flight Dynamics Principles},
   url = {http://ebookcentral.proquest.com/lib/brown/detail.action?docID=311429.},
   year = {2007}
}

@article{MaynardSmith1952,
   author = {John Maynard Smith},
   issue = {1},
   journal = {Evolution},
   pages = {127-129},
   title = {The Importance of the Nervous System in the Evolution of Animal Flight},
   volume = {6},
   year = {1952}
}

@article{Beatus2015,
   author = {Tsevi Beatus and John M. Guckenheimer and Itai Cohen},
   doi = {10.1098/rsif.2015.0075},
   issn = {17425662},
   issue = {105},
   journal = {Journal of the Royal Society Interface},
   month = {4},
   pmid = {25762650},
   publisher = {Royal Society of London},
   title = {Controlling roll perturbations in fruit flies},
   volume = {12},
   year = {2015}
}

@article{Hedrick2024,
   author = {Tyson L Hedrick and Emily Blandford and Haithem E Taha},
   doi = {10.1093/icb/icae076},
   issn = {1540-7063},
   journal = {Integrative And Comparative Biology},
   month = {6},
   pmid = {38897796},
   publisher = {Oxford University Press (OUP)},
   title = {Biomechanics of Insect Flight Stability and Perturbation Response},
   year = {2024}
}

@article{Breuer2022,
   author = {Kenneth Breuer and Mark Drela and Xiaozhou Fan and Matteo Di Luca},
   doi = {10.1007/s00348-022-03519-1},
   issn = {14321114},
   issue = {11},
   journal = {Experiments in Fluids},
   month = {11},
   publisher = {Springer Science and Business Media Deutschland GmbH},
   title = {Design and performance of an ultra-compact, low-speed, low turbulence level, wind tunnel for aerodynamic and animal flight experiments},
   volume = {63},
   year = {2022}
}

@article{Quinn2025,
   author = {Brooke L. Quinn and Jade L. Bajic and Santiago J. Romo and Ariel Wu and Alberto Bortoni and Kenneth Breuer and Sharon M. Swartz},
   doi = {10.1002/ar.25679},
   issn = {19328494},
   journal = {Anatomical Record},
   publisher = {John Wiley and Sons Inc},
   title = {Anatomical distribution and flight control function of wing sensory hairs in Seba's short-tailed bat},
   year = {2025}
}

@article{Rayner1995,
   author = {Jeremy M. V. Rayner and Coen Van Den Berg},
   issue = {8},
   journal = {Journal of Experimental Biology},
   month = {5},
   title = {The moment of inertia of bird wings and the inertial power requirement for flapping flight},
   volume = {198},
   year = {1995}
}

@article{TaylorSt2003,
   author = {Graham K. Taylor and Robert L. Nudds and Adrian L. R. Thomas},
   doi = {10.1038/nature02047},
   issn = {00280836},
   issue = {6959},
   journal = {Nature},
   month = {10},
   pages = {705-707},
   pmid = {14562100},
   title = {Flying and swimming animals cruise at a Strouhal number tuned for high power efficiency},
   volume = {425},
   year = {2003}
}

@article{Ellington1984,
   author = {C. P. Ellington},
   journal = {Philosophical Transactions of the Royal Society B},
   pages = {1-15},
   title = {THE AERODYNAMICS OF HOVERING INSECT FLIGHT. I. THE QUASI-STEADY ANALYSIS},
   volume = {305},
   url = {https://royalsocietypublishing.org/},
   year = {1984}
}

@article{Iriarte-Daz2011,
   author = {José Iriarte-Díaz and Daniel K. Riskin and David J. Willis and Kenneth S. Breuer and Sharon M. Swartz},
   doi = {10.1242/jeb.037804},
   issn = {00220949},
   issue = {9},
   journal = {Journal of Experimental Biology},
   month = {5},
   pages = {1546-1553},
   pmid = {21490262},
   title = {Whole-body kinematics of a fruit bat reveal the influence of wing inertia on body accelerations},
   volume = {214},
   year = {2011}
}

@article{Weis-Fogh1956,
   author = {T. Weis-Fogh and Martin Jensen},
   issue = {667},
   journal = {Philosophical Transactions of the Royal Society B},
   month = {7},
   title = {BIOLOGY AND PHYSICS OF LOCUST FLIGHT I. BASIC PRINCIPLES IN INSECT FLIGHT. A CRITICAL REVIEW},
   volume = {239},
   year = {1956}
}

@book{Nelson1998,
   author = {Robert C. Nelson},
   edition = {2},
   isbn = {0070462739},
   pages = {441},
   publisher = {WCB/McGraw Hill},
   title = {Flight stability and automatic control},
   year = {1998}
}

@article{Jones2006,
   author = {David E.H. Jones},
   doi = {10.1063/1.2364246},
   issn = {00319228},
   issue = {9},
   journal = {Physics Today},
   pages = {51-56},
   title = {The stability of the bicycle},
   volume = {59},
   year = {2006}
}

@book{Glauert1926,
   author = {Hermann Glauert},
   publisher = {Cambridge University Press},
   title = {The elements of aerofoil and airscrew theory},
   year = {1926}
}

@article{Cheney2022,
   author = {Jorn A. Cheney and Jeremy C. Rehm and Sharon M. Swartz and Kenneth S. Breuer},
   doi = {10.1242/jeb.243974},
   issn = {14779145},
   issue = {14},
   journal = {Journal of Experimental Biology},
   month = {7},
   pmid = {35762250},
   publisher = {Company of Biologists Ltd},
   title = {Bats actively modulate membrane compliance to control camber and reduce drag},
   volume = {225},
   year = {2022}
}

@article{Weston2026,
   author = {Kiran Weston and Huanglun Adam Zhu and Graham Keith Taylor and Christina Harvey},
   doi = {10.1098/rsif.2025.0868},
   issn = {17425662},
   issue = {236},
   journal = {Journal of the Royal Society, Interface},
   month = {3},
   pmid = {41784390},
   title = {Stability shifts in gliding flight: hawks morph from an unstable to stable state when navigating a gap},
   volume = {23},
   year = {2026}
}

@article{Ristroph2010,
   author = {Leif Ristroph and Attila J. Bergou and Gunnar Ristroph and Katherine Coumes and Gordon J. Berman and John Guckenheimer and Z. Jane Wang and Itai Cohen},
   doi = {10.1073/pnas.1000615107},
   issn = {00278424},
   issue = {11},
   journal = {Proceedings of the National Academy of Sciences of the United States of America},
   month = {3},
   pages = {4820-4824},
   pmid = {20194789},
   title = {Discovering the flight autostabilizer of fruit flies by inducing aerial stumbles},
   volume = {107},
   year = {2010}
}

\pagebreak

\onecolumn

\renewcommand{\thefigure}{S\arabic{figure}} 
\setcounter{figure}{0}


\subsection*{Quasi-static Assumption}

The quasi-static assumption (also known as the rigid body assumption or averaging approach) is frequently used in the flapping flight dynamics literature to simplify the analysis. When using the quasi-static assumption to evaluate how the aerodynamic forces and moments of the wingbeats impact the flight dynamics of the flier as a whole, we treat the unsteady aerodynamic forces and moments merely as averages over the wingbeat period. It requires that the body is nearly stationary in the time it takes to complete a wingbeat such that there is no coupling between the wingbeat and oscillations of the flier. The longitudinal motion of aircraft is typically defined by two rigid body modes: a short period mode and a longer `phugoid' mode \cite{Cook2007}. The quasi-static assumption is justified when the wingbeat frequency is much greater than the natural frequencies of these rigid body modes $f \gg \omega_n$ and the wing mass is much smaller than that of the body. Even if the wingbeat frequency and the highest natural frequency of the flier are of similar order, the quasi-static assumption may still be valid if the rigid body mode is highly damped. The quasi-static assumption is generally most important when evaluating dynamic stability since its validity determines whether the forces and moments during flapping need to be modelled as a function of state variables (speed and orientation) and time \cite{Taylor2005,Dietl2008,Ducci2021,Ducci2022} or just as a function of the state variables \cite{Taylor2003}. However, the quasi-static assumption also has the important role in static stability experiments of validating the use of tethered experiments as an approximation of real free flight. 

For birds, bats, and some large insects the condition $f \gg \omega_n$ is dubious \cite{Taylor2003} while for smaller insects it is generally safe \cite{Taha2015}, but more work is needed to assess typical natural frequencies of the rigid body longitudinal modes of animal fliers. Generally a ratio of $f / \omega_n > 10$ is assumed to be sufficient based on previous work with helicopters \cite{Taylor2002}.

As for the condition that the wing mass must be sufficiently small relative to the body, for insects this may seem reasonable since the total wing mass is less than 5\% of the body mass \cite{Taha2015,Orlowski2011} but for birds, bats, and similar scale ornithopters the total wing mass is closer to 10-15\% of the body mass \cite{Rayner1995,Grauer2012}. When the wings are this heavy, the center-of-mass (CoM) of the flier shifts substantially and the rotational inertia varies dramatically during a wingbeat. For example, in an ornithopter with a wingspan of 1.2 meters, it was estimated that the CoM traveled 10 cm and the moments of inertia varied by over 50\% during flapping \cite{Grauer2012}. Additionally, in a study of bat flight in a wind tunnel, the acceleration of the CoM estimated with an inertial model---including the moving wing mass---varied substantially from the acceleration of the CoM estimated with a fixed point on the body, with the two being completely out of phase at slower flight speeds \cite{Iriarte-Daz2011}.

\subsection*{Obtaining the Effective Wind Speed and Angle of Attack}

Depending on the angle of attack $\alpha$, the linear velocity of the wing can contribute in the drag or lift directions of the wing (see Figure \ref{schematic})

\begin{equation}
v_x = - r \dot{\theta} \sin \alpha, \qquad
v_n = - r \dot{\theta} \cos \alpha.
\end{equation}

The effective wind speed and angle of attack can be found through vector geometry:

\begin{equation}
\mathbf{c} = 
\begin{bmatrix}
-\cos(\alpha) \\ \sin(\alpha)
\end{bmatrix},
\qquad
\mathbf{u_{\text{eff}}} =
\begin{bmatrix}
v_x + U \\ v_n
\end{bmatrix},
\qquad
\alpha_{\text{eff}} = \sin^{-1} \left(\frac{\mathbf{c} \times \mathbf{u_{\text{eff}}}}{|\mathbf{u_{\text{eff}}}|}\right),
\end{equation}
where $\mathbf{c}$ is the unit vector defining the chord, $\mathbf{u_{\text{eff}}}$ describes the effective velocity vector of the wing's motion relative to the incoming air.

The effective wind speed and effective angle of attack are functions of time $t$, normalized spanwise position $\hat{r} = r/R$, angle of attack $\alpha$, Strouhal number $St$, and wingbeat frequency $f$:
\begin{equation}
    \frac{|\mathbf{u_{\text{eff}}}|}{U} = \sqrt{\pi ^2 \hat{r}^2 \text{St}^2 \sin ^2(2 \pi  f t) + 2 \pi  \hat{r} \text{St} \sin (\alpha ) \sin (2 \pi  f t)+1}
\end{equation}

\begin{equation}
    \alpha_{\text{eff}} = \sin ^{-1}\left(\frac{\sin (\alpha ) + \pi  \hat{r} \text{St} \sin (2 \pi  f t)}{\sqrt{\pi ^2 \hat{r}^2 \text{St}^2 \sin ^2(2 \pi  f t) + 2 \pi  \hat{r} \text{St} \sin (\alpha ) \sin (2 \pi  f t)+1}}\right)
\end{equation}

\pagebreak

\begin{figure}[htbp]
    \centering
    \includegraphics[width=0.5\textwidth]{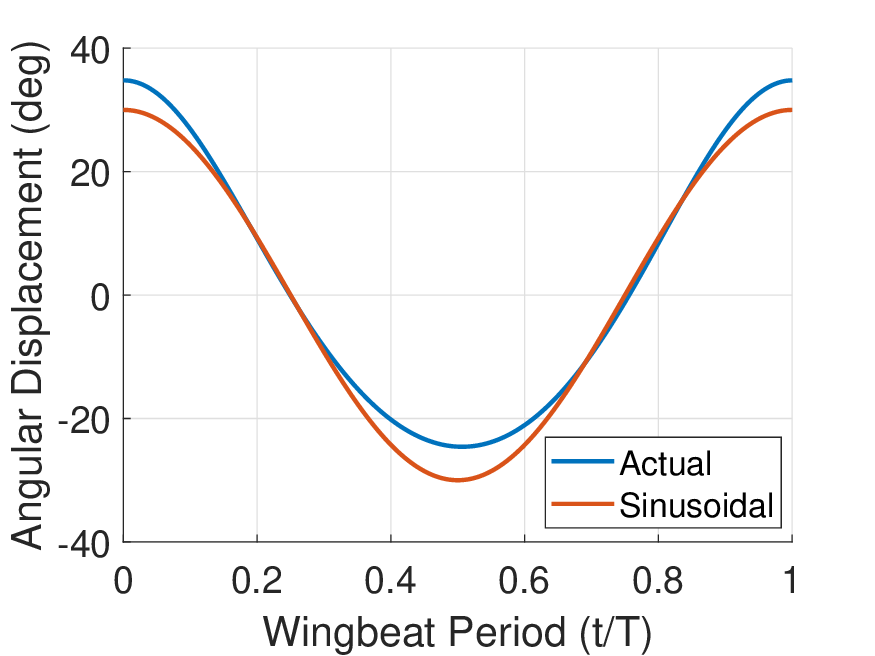}
    \caption{Actual kinematics determined from CAD model of robot flapping mechanism compared with sinusoidal kinematics used in QSBE model ($\theta = \frac{\pi}{6} \cos(2 \pi f t)$).}
    \label{true_vs_sin_kinematics}
\end{figure}

\begin{figure}[htbp]
    \centering
    \includegraphics[width=0.5\textwidth]{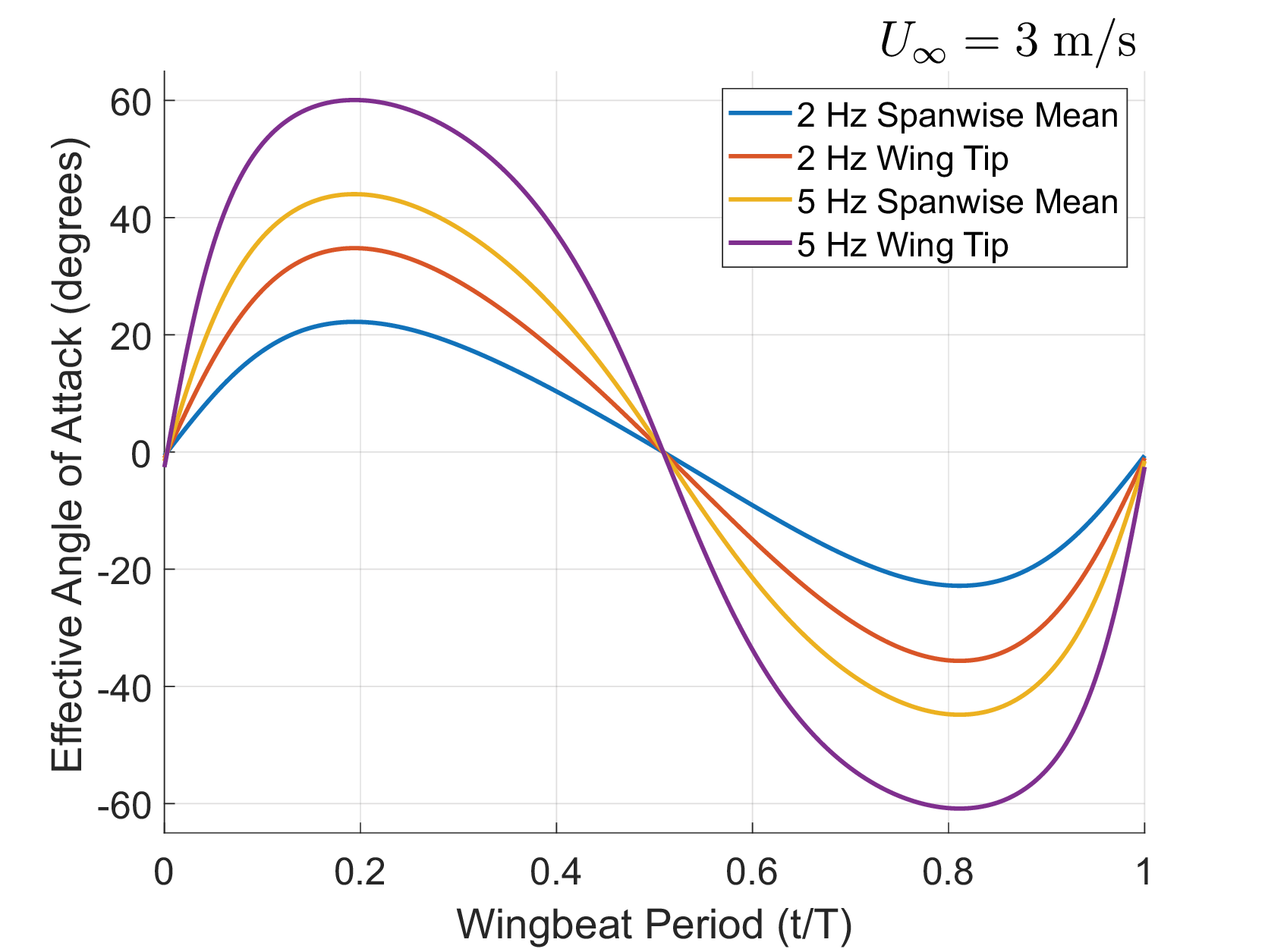}
    \caption{Variation of the effective angle of attack as a function of wingbeat frequency and spanwise location at the slowest wind speed tested (3 m/s). Considering all wind speeds, wingbeat frequencies, and spanwise positions; the purple line represents the most extreme variation of the angle of attack. ``Spanwise Mean" is the mean of the effective angle of attacks ranging from the wing root to the wing tip.}
    \label{eff_AoA}
\end{figure}

\begin{figure}[htbp]
    \centering
    \includegraphics[width=0.5\textwidth]{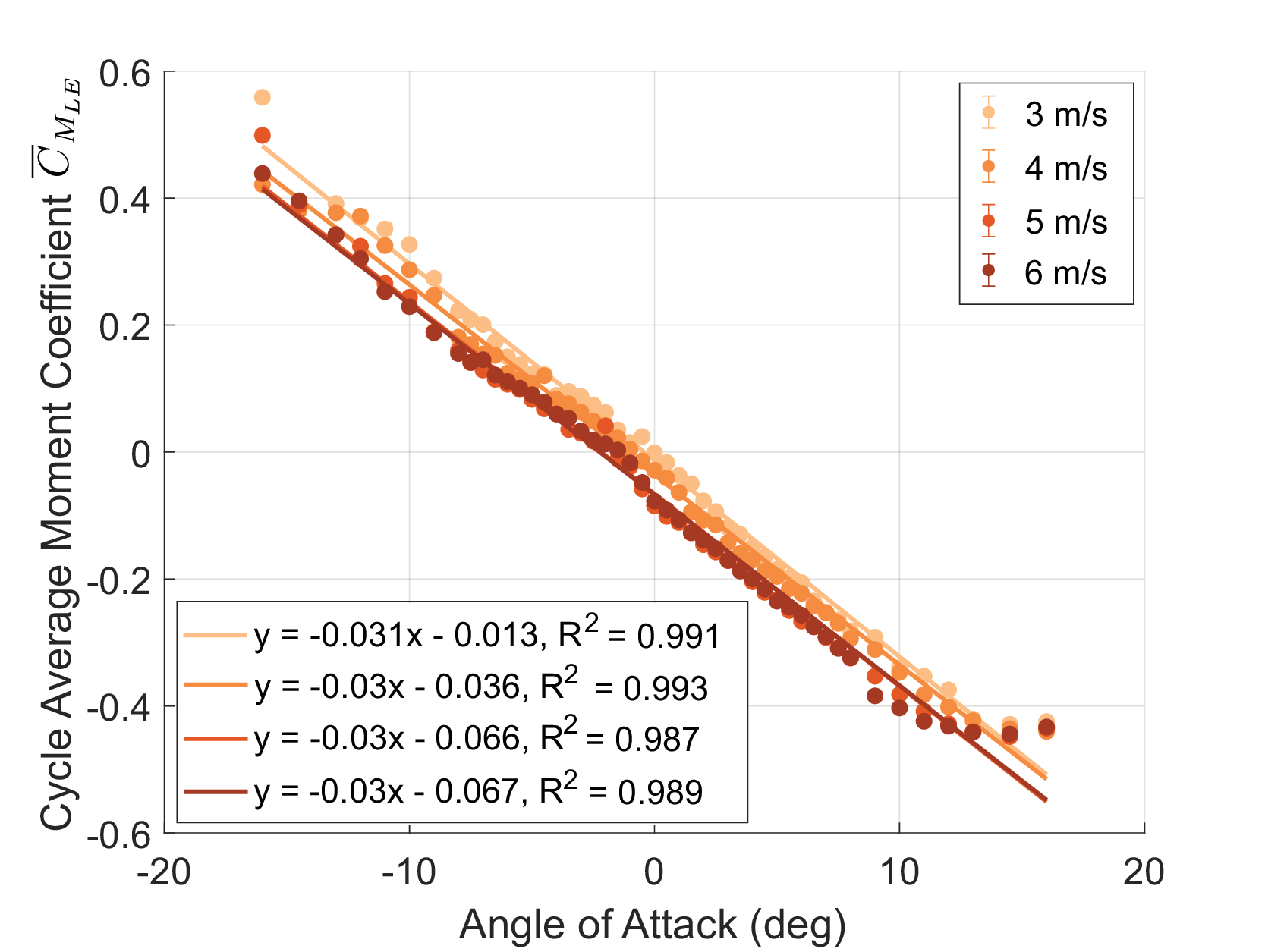}
    \caption{Cycle average pitch moment coefficient about the leading edge versus angle of attack for the robot in a static configuration (0 Hz) with wings fixed at midstroke for four different wind speeds. Pitch stiffness (stability slope) does not vary significantly for the range of Reynolds numbers tested.}
    \label{dMda_Re}
\end{figure}

\begin{figure*}[htbp]
    \centering
    \includegraphics[width=\textwidth]{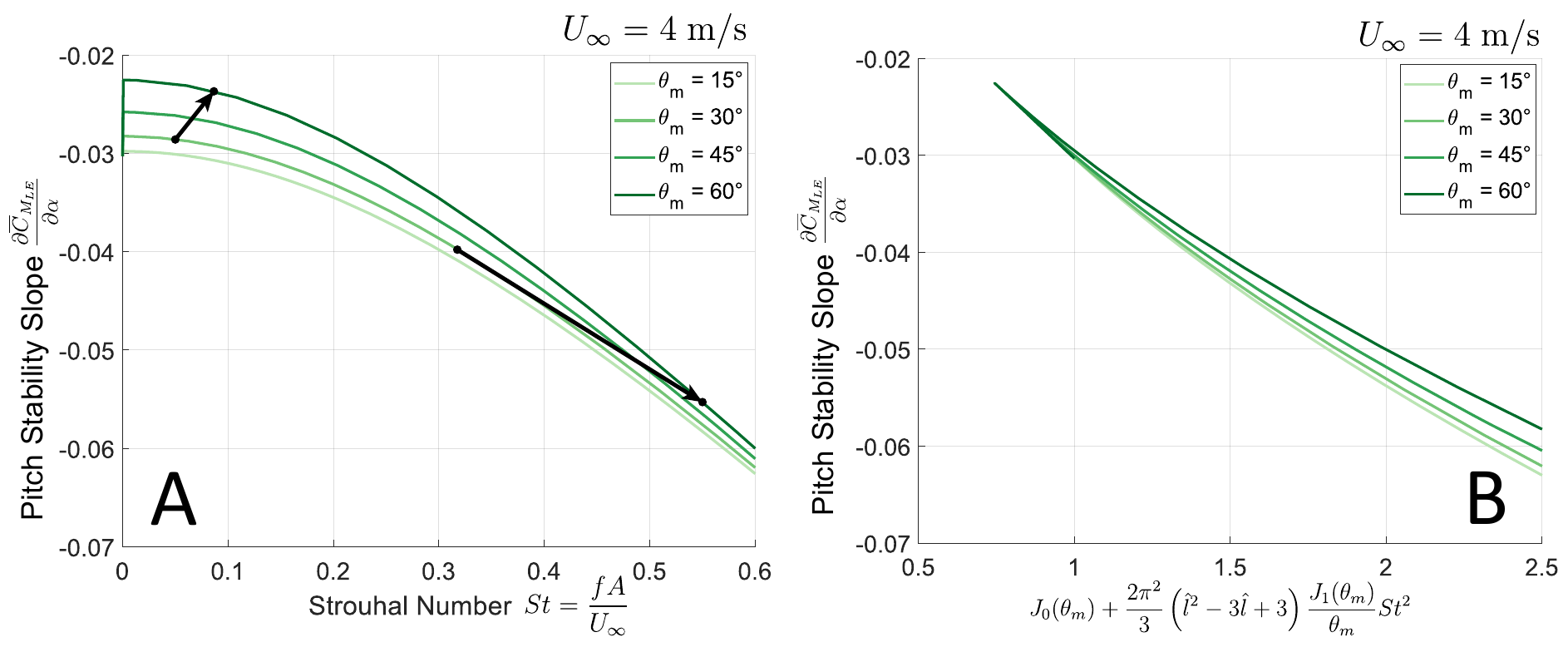}
    \caption{Effect of wingbeat amplitude and wingbeat frequency on pitch stiffness as predicted by the QSBE model. (a) uses a Strouhal scaling on the x-axis while (b) uses the scaling from the reduced QSBE model on the x-axis. The arrows in (a) show how the pitch stiffness changes when doubling the wingbeat amplitude while maintaining the same wingbeat frequency and wind speed. At low St, doubling the amplitude reduces the stiffness while at higher St it increases the stiffness. Note the failure of the Strouhal scaling to collapse the data. While the scaling in (b) certainly performs better than the Strouhal number, it is not perfect since only up to quadratic terms are retained from the model. The collapse is best in the limit of low St as expected since the series expansion is performed about the steady limit (St = 0).}
    \label{amp_model}
\end{figure*}

\begin{figure*}[htbp]
    \centering
    \begin{subfigure}[b]{0.5\textwidth}
        \centering
        \includegraphics[width=\textwidth]{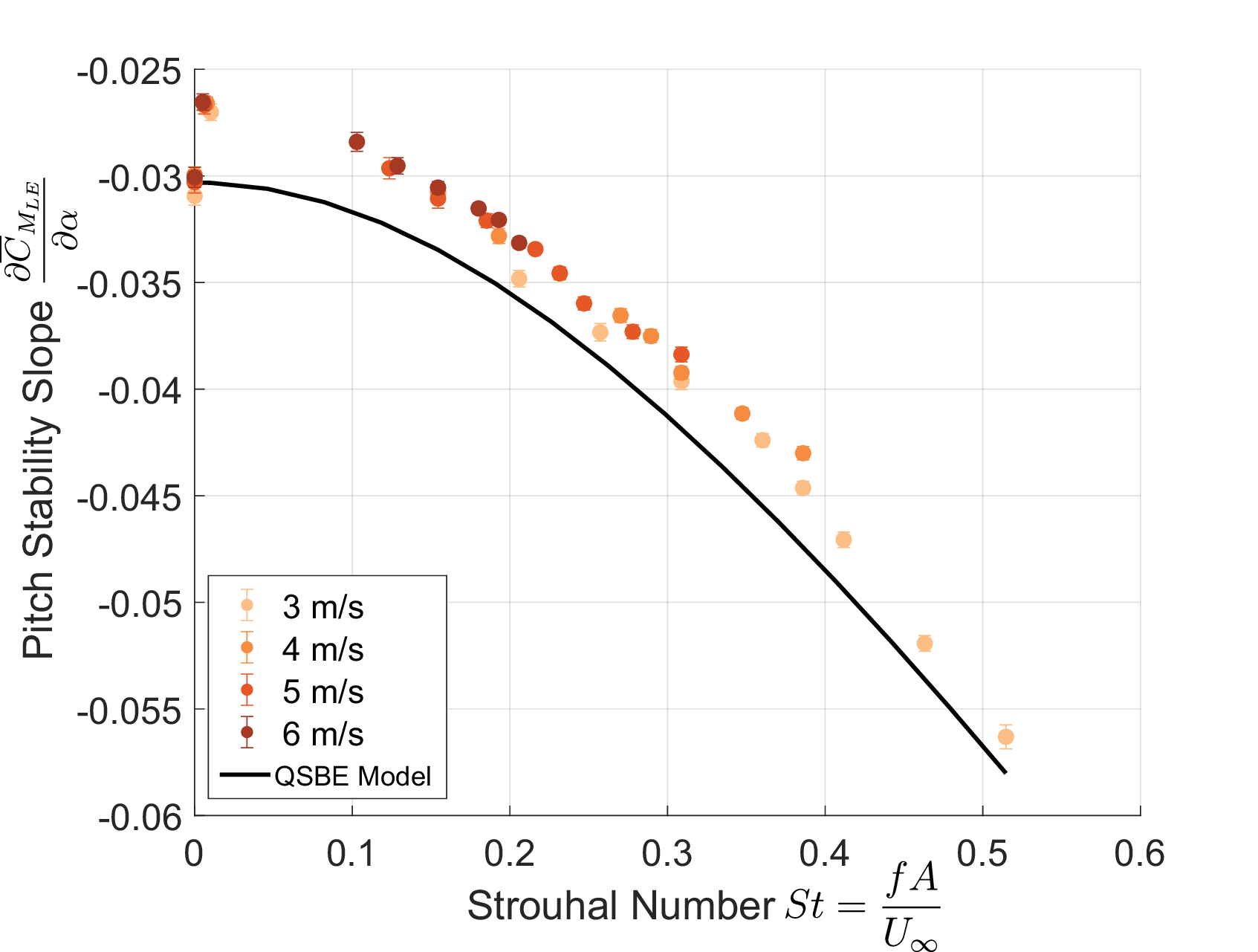}
        \caption{}
        \label{model_without_Area}
    \end{subfigure}
    \vspace{0.5em}
    \begin{subfigure}[b]{0.5\textwidth}
        \centering
        \includegraphics[width=\textwidth]{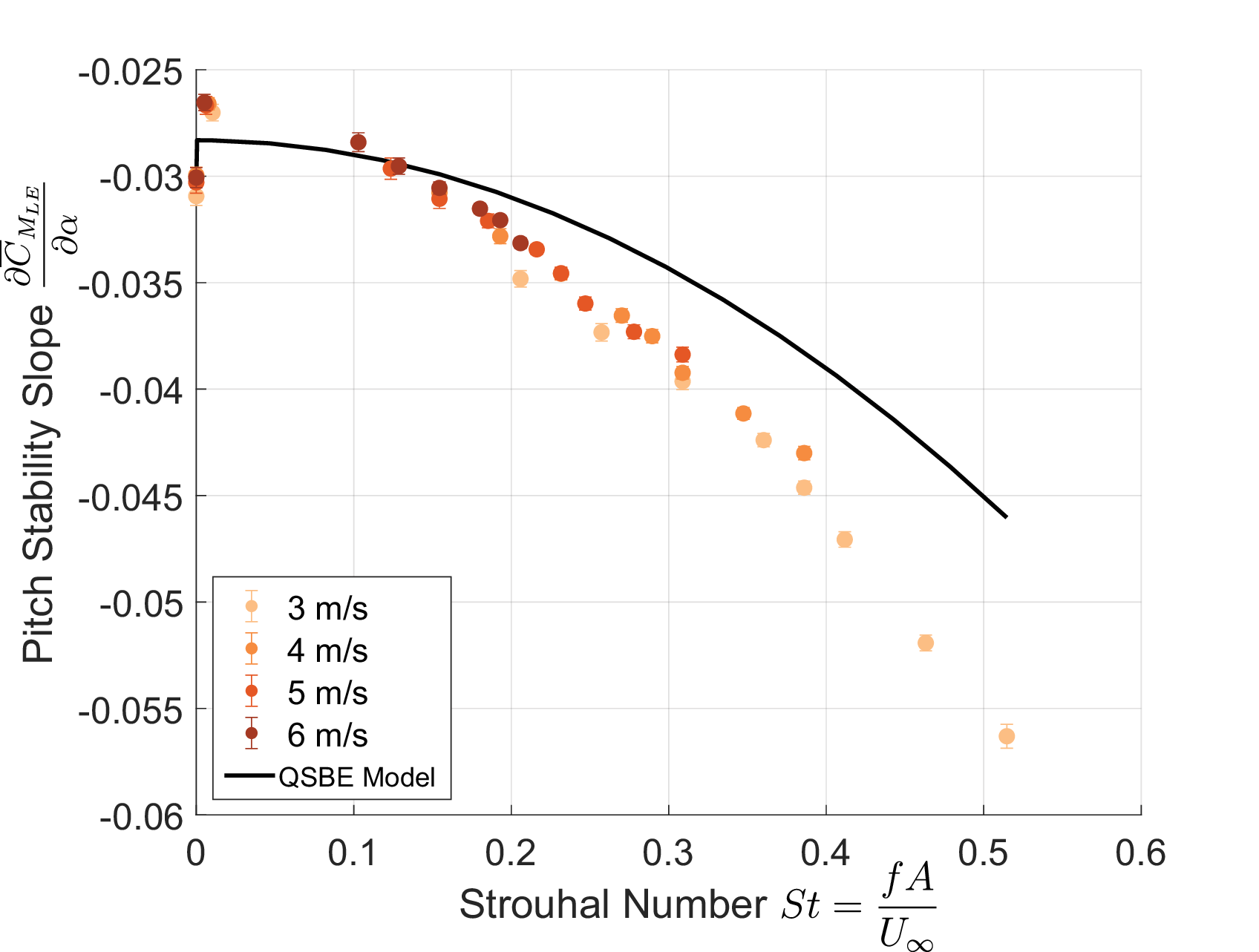}
        \caption{}
        \label{model_without_AoA}
    \end{subfigure}
    \vspace{0.5em}
    \begin{subfigure}[b]{0.5\textwidth}
        \centering
        \includegraphics[width=\textwidth]{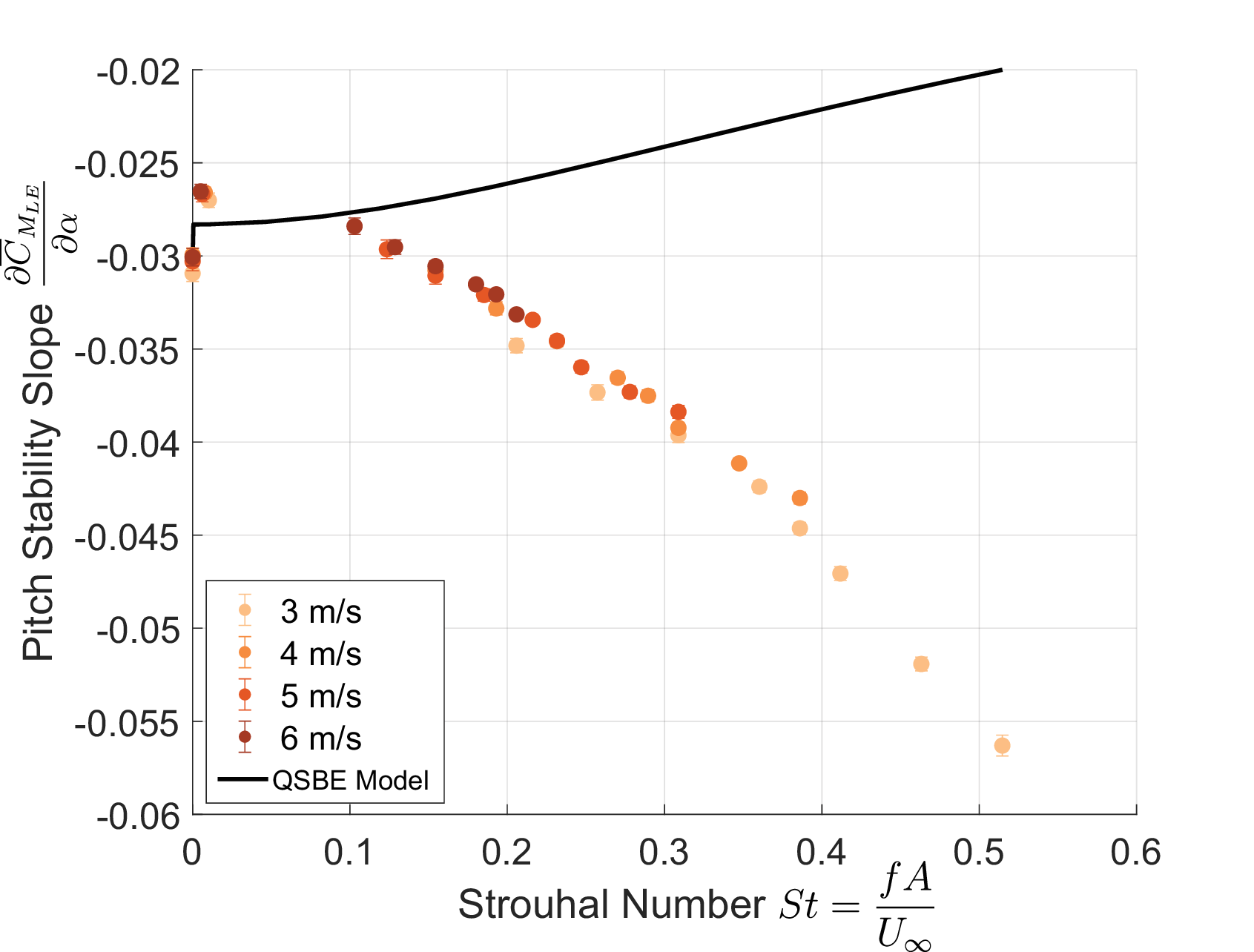}
        \caption{}
        \label{model_without_speed}
    \end{subfigure}
    \caption{Pitch stability slope (inverse pitch stiffness) as a function of Strouhal number with three different versions of the model: a) no $\cos\theta$ term, b) geometric angle of attack used rather than the effective angle of attack, c) no effective wind term. The agreement between the experimental data and the model is worst after removing the effective wind term, suggesting that this term dominates the model.}
    \label{model_variants}
\end{figure*}

\begin{figure}[htbp]
    \centering
    \includegraphics[width=0.6\textwidth]{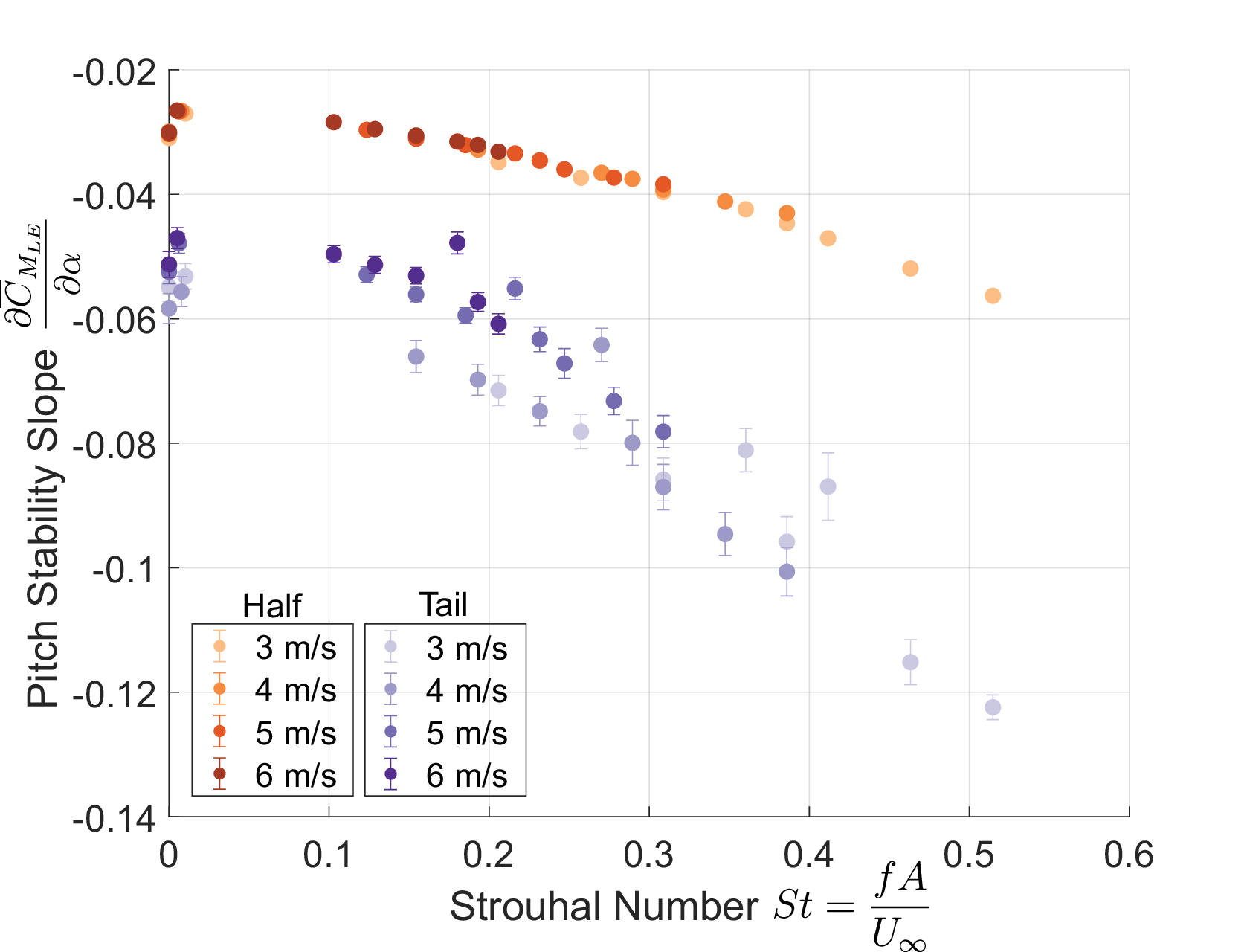}
    \caption{Pitch stability slope (inverse of pitch stiffness) as a function of Strouhal number for two different body configurations: ``half" and ``tail". The addition of the tail amplifies the stabilizing effect of flapping. Note that this data includes the body subtraction, so what's labeled as ``tail" is really describing the effect of the wing and its interaction with the tail rather than any contribution coming from the tail by itself.}
    \label{slopes_tail}
\end{figure}

\begin{figure*}[hbp]
    \centering
    \includegraphics[width=0.6\textwidth]{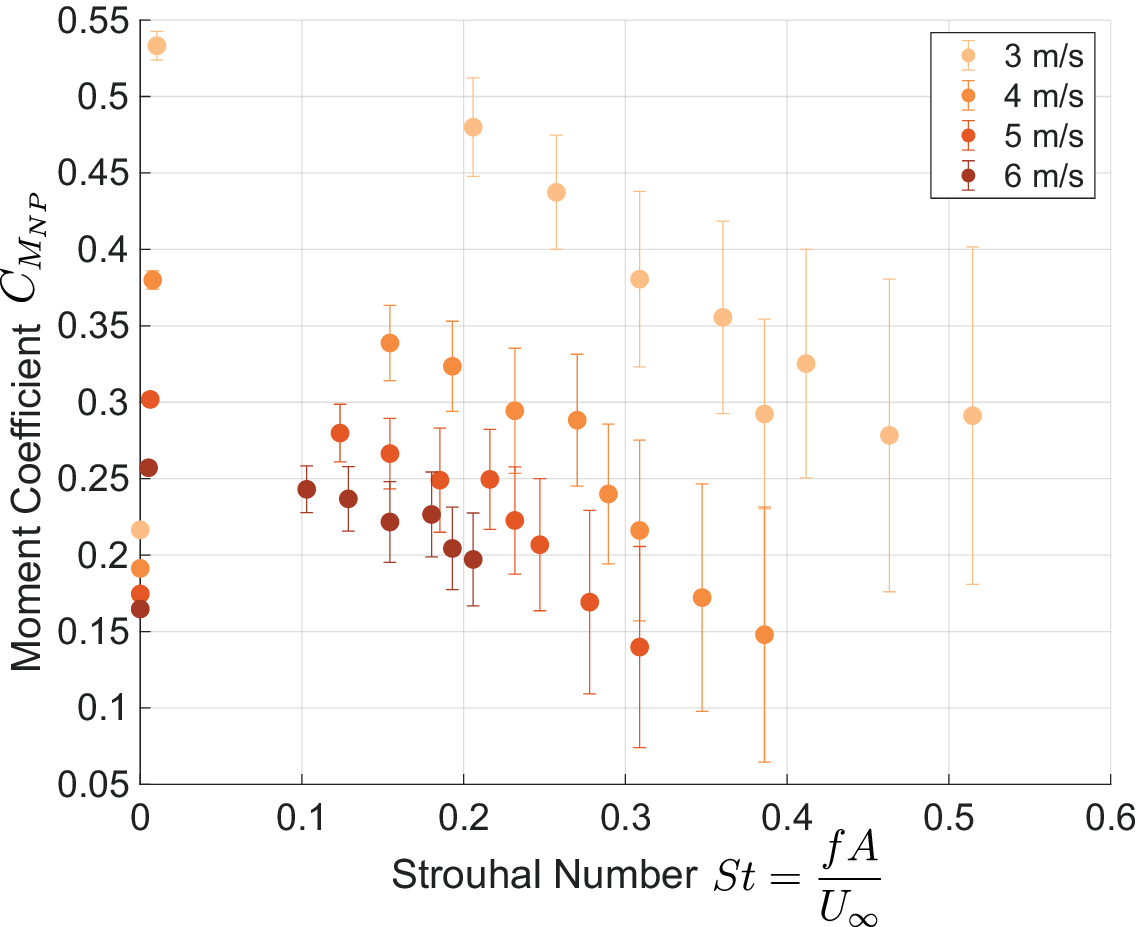}
    \caption{Pitch moment at the neutral point (NP) for the robot configuration with a tail and without the body subtraction. Since the moment about the NP is positive, the flier may be in equilibrium unlike Figure \ref{moment_NP}A. The estimated uncertainty is much higher for the ``tail'' configuration than the ``half'' configuration shown in the rest of the paper, presumably due to vibrations associated with the long body and tail.}
    \label{moment_NP_no_sub_tail}
\end{figure*}

\end{document}